\documentclass{article}
\pdfoutput=1
\usepackage{amsfonts,mathrsfs}
\usepackage{latexsym}
\usepackage{amssymb,amsbsy,amsmath,amsfonts,amssymb,amscd,amsthm}
\usepackage{subfigure}
\usepackage{graphicx}
\usepackage{hyperref}
\usepackage{dsfont}
\usepackage{mathtools}
\usepackage{pdfsync}
\usepackage{epsfig,epstopdf}
\usepackage{caption}
\usepackage{xcolor}
\usepackage{hyperref}
\usepackage{float}
\usepackage[applemac]{inputenc}
\usepackage{color} 
\usepackage{bbm} 

\newtheorem{theorem}{Theorem}

\newtheorem{proposition}{Proposition}
\newtheorem{remark}{Remark}

\newcommand {\Chi} {{\bf \raise 2pt \hbox{$\chi$}} }

\newcommand{\argmax}[1]{\underset{#1}{\operatorname{arg}\,\operatorname{max}}\;}
\newcommand{\argmin}[1]{\underset{#1}{\operatorname{arg}\,\operatorname{min}}\;}

\newcommand{\beq}{\begin{equation}}
\newcommand{\eeq}{\end{equation}}
\newcommand{\beqa}{\begin{eqnarray}}
\newcommand{\eeqa}{\end{eqnarray}}
\newcommand{\bea} {\begin{array}{rl}}
\newcommand{\eea} {\end{array}}
\newcommand{\bepa}{\left\{ \begin{array}{l}}
\newcommand{\eepa} {\end{array}\right.}

\title{Evolutionary dynamics in vascularised tumours under chemotherapy}

\author{Chiara Villa\thanks{University of St Andrews, UK 
  (cv23@st-andrews.ac.uk)} 
\and
Mark A. J. Chaplain\thanks{University of St Andrews, UK 
  (majc@st-andrews.ac.uk)}
\and
Tommaso Lorenzi\thanks{University of St Andrews, UK 
  (tl47@st-andrews.ac.uk)}
}

\begin{document}
\date{}
\maketitle

\begin{abstract}
We consider a mathematical model for the evolutionary dynamics of tumour cells in vascularised tumours under chemotherapy. The model comprises a system of coupled partial integro-differential equations for the phenotypic distribution of tumour cells, the concentration of oxygen and the concentration of a chemotherapeutic agent. In order to disentangle the impact of different evolutionary parameters on the emergence of intra-tumour phenotypic heterogeneity and the development of resistance to chemotherapy, we construct explicit solutions to the equation for the phenotypic distribution of tumour cells and provide a detailed quantitative characterisation of the long-time asymptotic behaviour of such solutions. Analytical results are integrated with numerical simulations of a calibrated version of the model based on biologically consistent parameter values. The results obtained provide a theoretical explanation for the observation that the phenotypic properties of tumour cells in vascularised tumours vary with the distance from the blood vessels. Moreover, we demonstrate that lower oxygen levels may correlate with higher levels of phenotypic variability, which suggests that the presence of hypoxic regions supports intra-tumour phenotypic heterogeneity. Finally, the results of our analysis put on a rigorous mathematical basis the idea, previously suggested by formal asymptotic results and numerical simulations, that hypoxia favours the selection for chemoresistant phenotypic variants prior to treatment. Consequently, this facilitates
the development of resistance following chemotherapy.
\end{abstract}

\section{Introduction}
\label{intro}
Previous empirical and theoretical work has suggested that spatial variation in oxygen levels can foster the emergence of intra-tumour phenotypic heterogeneity~\cite{alfarouk2013riparian,gallaher2013evolution,gay2016tumour,gillies2012evolutionary,loeb2001mutator,marusyk2012intra,molavian2009fingerprint,sun2015intra}. In particular, it has been hypothesised that the nonlinear interplay between impaired oxygen delivery caused by structural abnormalities present in the tumour vasculature~\cite{dewhirst2008cycling,fukumura2010tumor,jain1988determinants,jordan2012targeting,padhani2007imaging,vartanian2014gbm,vaupel1989blood}, limited oxygen diffusion and oxygen consumption by tumour cells may lead to the creation of distinct ecological niches in vascularised tumours, whereby tumour cells with different phenotypic characteristics can be selected~\cite{alfarouk2013riparian,casciari1992variations,gatenby2007cellular,hockel2001tumor,ibrahim2017defining,lloyd2016darwinian}. This hypothesis is supported by a growing body of experimental and clinical studies indicating that: well-oxygenated parts of the tumour are densely populated by cells characterised by higher oxygen uptake and faster proliferation via aerobic pathways; hypoxic parts of the tumour (\emph{i.e.} regions where oxygen levels are below normal physiological levels) are mainly occupied by cells that display higher levels of hypoxia-inducible factors, such as HIF-1~\cite{dewhirst2008cycling,giatromanolaki2001relation,lee2004hypoxia,padhani2007imaging,robey2005hypoxia,semenza2003targeting,strese2013effects,tannock1968relation,zhao2013targeting}, which typically correlate with slower proliferation via anaerobic pathways and higher levels of resistance to chemotherapy~\cite{brown1998unique, durand1998identification}. 

In this paper, we use a mathematical model for the evolutionary dynamics of tumour cells in vascularised tumours under chemotherapy to gain a deeper understanding of the adaptive process that underpins the emergence of intra-tumour phenotypic heterogeneity and the development of resistance to chemotherapeutic agents. The model comprises a system of coupled partial integro-differential equations for the phenotypic distribution of tumour cells, the concentration of oxygen and the concentration of a chemotherapeutic agent. 

Non-local partial differential equations (PDEs) similar to the one that governs the evolution of the phenotypic distribution of tumour cells in our model and related integro-differential equations have recently received increasing attention from the mathematical community -- see for instance ~\cite{alfaro2017effect,alfaro2013travelling,arnold2012existence,bouin2014travelling,bouin2012invasion,bouin2015hamilton,calvez2018non,domschke2017structured,hodgkinson2018signal,jabin2016selection,mirrahimi2015asymptotic}. In particular, our work follows earlier papers on the analysis and numerical simulation of integro-differential equations and non-local PDEs modelling the emergence of intra-tumour phenotypic heterogeneity~\cite{jabin2016selection,lorenzi2018role,lorz2015modeling,mirrahimi2015asymptotic,villa2019modelling}. The focus of these papers is on the cases where tumour cells do not change their phenotypic state or the rate of phenotypic variation is small. The main novelty of our work is that we allow tumour cells to undergo spontaneous epimutations (\emph{i.e.} heritable phenotypic changes that occur randomly due to non-genetic instability and are not induced by any selective pressure~\cite{huang2013genetic}) and we do not impose any smallness assumptions on the rate at which such phenotypic changes occur. In this more general scenario, building upon the method of proof presented in~\cite{almeida2019evolution,ardavseva2020evolutionary,chisholm2016evolutionary,lorenzi2015dissecting}, we carry out an analytical study of evolutionary dynamics. In particular, we construct explicit solutions to the equation for the phenotypic distribution of tumour cells and, considering the case where the concentrations of oxygen and chemotherapeutic agent are stationary, we provide a detailed quantitative characterisation of the long-time asymptotic behaviour of such solutions. The analytical results obtained are integrated with numerical simulations of a calibrated version of the model based on biologically consistent parameter values, in order to further assess the impact of the dynamics of oxygen and chemotherapeutic agent on the phenotypic evolution of tumour cells. 

The paper is organised as follows. In Section~\ref{sec:model}, we introduce the equations of the model and the underlying modelling assumptions. In Section~\ref{sec:analysis}, we present the results of our analytical study of evolutionary dynamics. In Section~\ref{sec:numsol}, we report on numerical solutions that confirm and extend the analytical results obtained. Section~\ref{sec:disc} concludes the paper and provides a brief overview of possible research perspectives.

\section{Mathematical model}
\label{sec:model}
We model the evolution of tumour cells within a region of a vascularised tumour along with the dynamical interactions that occur between tumour cells and both oxygen and a chemotherapeutic agent, which are released from the intra-tumoural vascular network. 

The tumour region is approximated as a bounded set $\Omega \subset \mathbb{R}^d$, with smooth boundary $\partial \Omega$, where $d=1,2,3$ depending on the biological scenario under study. The spatial position of tumour cells is described by a vector ${\bf x} \in \overline{\Omega}$ and the phenotypic state of every cell is modelled by a scalar variable $y \in \mathbb{R}$, which represents the rescaled level of a hypoxia-inducible factor. Building upon the ideas presented in~\cite{lorenzi2016tracking,pisco2015non}, we assume that there is a sufficiently high level of expression of the hypoxia-inducible factor $y^H$ conferring both the highest rate of cellular division via anaerobic energy pathways and the highest level of resistance to chemotherapy, while there is a sufficiently low level of expression of the hypoxia-inducible factor $y^L < y^H$ providing the highest rate of cellular division via aerobic energy pathways. Without loss of generality, we define $y^H:=1$ and $y^L:=0$, so that values of $y \to 1$ correspond to phenotypic variants with higher rates of cellular division via anaerobic energy pathways and higher levels of chemoresistance (\emph{i.e.} anaerobic and chemoresistant phenotypic variants), whereas values of $y \to 0$ correspond to phenotypic variants with higher rates of cellular division via aerobic energy pathways (\emph{i.e.} aerobic phenotypic variants).

The phenotypic distribution of tumour cells at time $t \in [0,\infty)$ and position ${\bf x}$ is described by the function $n(t,{\bf x},y)$, while the functions $s(t,{\bf x})$ and $c(t,{\bf x})$ describe, respectively, the oxygen concentration and the concentration of the chemotherapeutic agent at time $t$ and position ${\bf x}$. Moreover, at each time $t$, we define the density of tumour cells at position ${\bf x}$ as
\begin{equation}\label{rho}
\rho(t,{\bf x}) := \int_{\mathbb{R}}  n(t,{\bf x},y) \, {\rm d}y, 
\end{equation}
while the local mean phenotypic state and the related variance are defined, respectively, as
\begin{equation}\label{musigma}
\mu(t,{\bf x}) := \frac{1}{\rho(t,{\bf x})} \int_{\mathbb{R}} y \, n(t,{\bf x},y) \, {\rm d}y \quad \text{and} \quad \sigma^2(t,{\bf x}) := \frac{1}{\rho(t,{\bf x})} \int_{\mathbb{R}} y^2 \, n(t,{\bf x},y) \, {\rm d}y - \mu^2(t,{\bf x}).
\end{equation}

\subsection{\bf Dynamics of tumour cells}
The local phenotypic distribution of tumour cells $n(t,{\bf x},y)$ is governed by the following non-local PDE
\beq
\label{eq:n}
\begin{cases}
\displaystyle{\partial_{t} n = \beta \, \partial^2_{yy} n + R\big(y,\rho(t,{\bf x}),s(t,{\bf x}), c(t,{\bf x})\big) n}, \quad (t,{\bf x},y) \in (0,\infty) \times \overline{\Omega} \times \mathbb{R},
\\\\
\displaystyle{\rho(t,{\bf x}) :=  \int_{\mathbb{R}} n(t,{\bf x},y) \, {\rm d}y}.
\end{cases}
\eeq
In the reaction-diffusion equation~\eqref{eq:n}, the diffusion term models the effect of spontaneous epimutations, which occur at rate $\beta>0$~\cite{chisholm2015emergence, lorenzi2016tracking}, while the non-local reaction term models the effect of cell division and death. The function $R\big(y,\rho(t,{\bf x}),s(t,{\bf x}),c(t,{\bf x})\big)$ represents the fitness of tumour cells in the phenotypic state $y$ at position ${\bf x}$ and time $t$ under the local environmental conditions given by the cell density $\rho(t,{\bf x})$, the oxygen concentration $s(t,{\bf x})$ and the  concentration of chemotherapeutic agent $c(t,{\bf x})$ (\emph{i.e.} the phenotypic fitness landscape of the tumour at position ${\bf x}$ and time $t$). In particular, we consider
\begin{equation}\label{def:R}
R\big(y,\rho,s,c\big) :=  p(y,s) - \zeta \, \rho - k(y,c) 
\end{equation}
with
\begin{equation}\label{def:p}
p(y,s) := f(y) + g(y,s).
\end{equation}
Here, $\zeta > 0$, $f(y)$ is a ${\rm C}^2$-function such that
\beq
\argmax{y \in \mathbb{R}} f(y) = 1, \quad f(1) > 0, \quad \partial^2_{yy} f < 0, 
\label{assf}
\eeq
$g(y,s)$ is a ${\rm C}^2$-function of $y$ and a ${\rm C}^1$-function of $s$ that satisfies the following assumptions
\beq
\argmax{y \in \mathbb{R}} g(y,s) = 0, \quad g(0,s) > 0, \quad \partial^2_{yy} g(\cdot,s) < 0 \quad \forall \, s \in (0,\infty), \quad \lim_{s \to \infty} g(0,s) > f(1)
\label{assg1}
\eeq
\beq
g(\cdot,0) = 0, \quad \partial_s |g(\cdot,s)| \geq 0 \quad \forall \, s \in (0,\infty),
\label{assg2}
\eeq
and $k(y,c)$ is a ${\rm C}^2$-function of $y$ and a ${\rm C}^1$-function of $c$ that satisfies the following assumptions
\beq
\argmin{y \in \mathbb{R}} k(y,c) = 1, \quad k(1,c) = 0,  \quad \partial^2_{yy} k(\cdot,c) >0 \quad \forall \, c \in (0,\infty), 
\label{assk1}
\eeq
\beq
k(\cdot,0) = 0, \quad \partial_c k(\cdot,c) \geq 0 \quad \forall \, c \in (0,\infty).
\label{assk2}
\eeq
Definition~\eqref{def:R} along with assumptions~\eqref{assk1} and~\eqref{assk2} models a biological scenario whereby the background fitness of tumour cells in the phenotypic state $y$ at position ${\bf x}$ and time $t$ is given by a function $p(y,s(t,{\bf x}))$, the value of which is reduced:
\begin{itemize}
\item  due to competition for limited space, by a certain amount which is the same for all phenotypic variants and is proportional to $\rho(t,{\bf x})$, with a proportionality constant $\zeta$ that is related to the local carrying capacity of the tumour;
\item[]
\item  due to the cytotoxic action of the chemotherapeutic agent, by a certain amount $k(y,c)$ which increases monotonically with the concentration of the chemotherapeutic agent $c$ and is smaller for phenotypic variants with $y \to 1$, which are characterised by higher levels of chemoresistance, and is null for the phenotypic variant corresponding to $y=1$, since such a phenotypic variant is assumed to be completely resistant to the chemotherapeutic agent.
\end{itemize}

Definition~\eqref{def:p} corresponds to the case where the background fitness $p(y,s)$ is defined as a linear combination of the background fitness associated with anaerobic energy pathways $f(y)$ and the background fitness associated with aerobic energy pathways $g(y,s)$. In particular, assumptions~\eqref{assf}-\eqref{assg2} translate into mathematical terms the following biological ideas:
\begin{itemize}
\item The state $y=1$ corresponds to the phenotypic variant with the maximal background fitness associated with anaerobic energy pathways, whereas the  state $y=0$ corresponds to the phenotypic variant with the maximal background fitness associated with aerobic energy pathways.
\item[]
\item Due to the fact that less fit phenotypic variants are driven to extinction by natural selection, the background fitness associated with anaerobic (or aerobic) energy pathways can be negative for phenotypic variants with values of $y$ sufficiently far from $1$ (or $0$).
\item[]
\item Because of the fitness cost associated with a less efficient anaerobic metabolism~\cite{basanta2008evolutionary}, the maximal background fitness of aerobic phenotypic variants in well-oxygenated environments is larger than the maximal background fitness of anaerobic phenotypic variants. 
\item[]
\item In the absence of oxygen, the background fitness $p(y,s)$ coincides with the background fitness associated with anaerobic energy pathways $f(y)$.
\item[]
\item The larger is the oxygen concentration, the stronger is the impact of the background fitness associated with aerobic energy pathways $g(y,s)$ on the background fitness $p(y,s)$.
\end{itemize}

In particular, following the modelling strategies presented in~\cite{lorenzi2018role}, here we use the definitions
\begin{equation}\label{def:fgk}
f(y) := \varphi \; \Big[1 - (1-y)^2\Big], \quad g(y,s) := \gamma_s \frac{s}{\alpha_s + s} \; \big(1-y^2\big), \quad k(y,c) :=  \gamma_c \frac{c}{\alpha_c + c} \; (1-y)^2,
\end{equation}
where $\varphi>0$ is the maximal background fitness of anaerobic phenotypic variants, $\gamma_s>\varphi$ is the maximal background fitness of aerobic phenotypic variants, $\alpha_s>0$ and $\alpha_c>0$ are the Michaelis-Menten constants of oxygen and chemotherapeutic agent respectively, and $\gamma_c>0$ is the maximal reduction of the background fitness of aerobic phenotypic variants due to the cytotoxic action of the chemotherapeutic agent. Definitions~\eqref{def:fgk} satisfy assumptions~\eqref{assf}-\eqref{assk2}, ensure analytical tractability of the model and lead to a fitness function $R\big(y,\rho,s,c\big)$ that is close to the approximate fitness landscapes which can be inferred from experimental data through regression techniques -- see, for instance, equation (1) in~\cite{otwinowski2014inferring}. In fact, with these definitions, after a little algebra, the difference $p(y,s) - k(y,c)$ in~\eqref{def:R} can be rewritten as
\begin{equation}\label{pn2ori}
p(y,s) - k(y,c) = a(s,c) - b(s,c) \left(y - h(s,c)\right)^2
\end{equation}
where
\beq
\label{eq:a} 
a(s,c) := \gamma_s \dfrac{s}{\alpha_s + s} - \gamma_c \dfrac{c}{\alpha_c + c} + \frac{\left(\varphi + \gamma_c \dfrac{c}{\alpha_c + c} \right)^2}{\varphi+\gamma_s \dfrac{s}{\alpha_s + s} + \gamma_c \dfrac{c}{\alpha_c + c}},
\eeq
\beq
\label{eq:b} 
b(s,c) := \varphi + \gamma_s \dfrac{s}{\alpha_s + s} + \gamma_c \dfrac{c}{\alpha_c + c}
\eeq
and
\beq
\label{eq:h} 
h(s,c) := \frac{\varphi + \gamma_c \dfrac{c}{\alpha_c + c}}{\varphi+\gamma_s \dfrac{s}{\alpha_s + s} + \gamma_c \dfrac{c}{\alpha_c + c}}.
\eeq
Here, $a(s,c)$ is the maximum fitness, $h(s,c)$ is the fittest phenotypic state and $b(s,c)$ is the selection gradient under the environmental conditions corresponding to the oxygen concentration $s(t,{\bf x})$ and the concentration of chemotherapeutic agent $c(t,{\bf x})$. We remark that $b(s,c)$ is a selection gradient in that it provides a measure of the strength of the selective pressure exerted on tumour cells by oxygen and the chemotherapeutic agent~\cite{lande1983measurement}. Notice that, 
$$
h : [0,\infty) \times [0,\infty) \to [0,1], \qquad \lim_{s\to0} h(s,\cdot) = 1, \qquad \lim_{s\to\infty} h(s,0) = \frac{1}{1+ \dfrac{\gamma_s}{\varphi}},
$$
and
$$
\lim_{c\to\infty} h(s,c) = \dfrac{1}{1+ \dfrac{\gamma_s}{\varphi + \gamma_c} \dfrac{s}{\alpha_s + s}} \quad \forall s \in [0,\infty).
$$
Hence, consistent with our modelling assumptions, 
\begin{itemize}
\item for any concentrations of oxygen and chemotherapeutic agent, the fittest phenotypic state is between $y=0$ (\emph{i.e.} the state corresponding to the phenotypic variant with the highest rate of cellular division via aerobic energy pathways) and $y=1$ (\emph{i.e.} the state corresponding to the phenotypic variant with the highest rate of cellular division via anaerobic energy pathways and the highest level of resistance to chemotherapy);
\item[]
\item in hypoxic conditions ({\it i.e.} when $s\to0$), the fittest phenotypic state is $y=1$;
\item[]
\item when there is no chemotherapeutic agent ({\it i.e.} when $c\equiv0$), in well-oxygenated environments ({\it i.e.} when $s\to\infty$) the larger is the ratio between the maximal background fitness of aerobic phenotypic variants $\gamma_s$ and the maximal background fitness of anaerobic phenotypic variants $\varphi$, the closer the fittest phenotypic state will be to $y=0$;
\item[]
\item under high-dose chemotherapy, the smaller is the ratio between the maximal background fitness of aerobic phenotypic variants $\gamma_s$ and the maximal reduction of the background fitness of aerobic phenotypic variants due to the cytotoxic action of the chemotherapeutic agent $\gamma_c$, the closer the fittest phenotypic state will be to $y=1$.
\end{itemize}

\subsection{\bf Dynamics of abiotic factors}
The oxygen concentration $s(t,{\bf x})$ and the concentration of chemotherapeutic agent $c(t,{\bf x})$ are governed by the following partial integro-differential equations
\beq
\displaystyle{\partial_{t}s = D_s \, \Delta_{\bf x} s - \int_{\mathbb{R}} r_s(y,s)  \, n(t,{\bf x},y) \, {\rm d}y - \lambda_s s + q_s(t,{\bf x}),} \quad (t,{\bf x}) \in (0,\infty) \times \Omega
\label{eq:s}
\eeq
and
\beq
\displaystyle{\partial_{t}c = D_c \, \Delta_{\bf x} c - \int_{\mathbb{R}} r_c(y,c)  \, n(t,{\bf x},y) \, {\rm d}y - \lambda_c c + q_c(t,{\bf x}),} \quad (t,{\bf x}) \in (0,\infty) \times \Omega
\label{eq:c}
\eeq
coupled with~\eqref{eq:n} and subject to zero-flux boundary conditions, \emph{i.e.}
\beq
\label{eq:bcs}
\nabla_{\bf x} s \cdot {\bf u} = 0 \quad \text{and} \quad \nabla_{\bf x} c \cdot {\bf u} = 0 \quad \text{on }  \partial\Omega,
\eeq
where ${\bf u}$ is the unit normal to $\partial\Omega$ that points outward from $\Omega$. In~\eqref{eq:s} and~\eqref{eq:c}, the parameters $D_s>0$ and $D_c>0$ are the diffusion coefficients of oxygen and chemotherapeutic agent, the functions $r_s(y,s)$ and $r_c(y,c)$ are the consumption rates of oxygen and chemotherapeutic agent by tumour cells in the phenotypic state $y$, the parameters $\lambda_s>0$ and $\lambda_c>0$ are the natural decay rates of oxygen and chemotherapeutic agent, and the source terms $q_s(t,{\bf x})$ and $q_c(t,{\bf x})$ model the influx of oxygen and chemotherapeutic agent from the intra-tumoural blood vessels at position ${\bf x} \in \Omega$ and at time $t$.

We assume that the oxygen is consumed only by phenotypic variants corresponding to values of $y$ for which the fitness associated with aerobic energy pathways $g(y,s)$ is positive and we let oxygen consumption occur at a rate proportional to $g(y,s)$. Moreover, we assume that the chemotherapeutic agent is consumed by phenotypic variants corresponding to different $y$ at different rates proportional to the amount $k(y,c)$ by which their background fitness is reduced due to the cytotoxic action of the chemotherapeutic agent. In accordance with these assumptions, we use the following definitions
\beq
\label{eq:rsrc}
r_s(y,s) := \eta_s \, \left(g(y,s)\right)_+ \quad \text{and} \quad r_c(y,s) := \eta_c \, k(y,c),
\eeq
where $\eta_s>0$ and $\eta_c>0$ are constants of proportionality and $\left(\cdot \right)_+$ denotes the positive part of $\left(\cdot \right)$. As done in~\cite{villa2019modelling}, we  let $\omega \subset \Omega$ be the set of points within the tumour tissue which are occupied by blood vessels and, since we do not consider the formation of new blood vessels, we assume $\omega$ to be given and remain constant in time. Therefore, we define the source terms $q_s$ and $q_c$ as
\beq
\label{q}
q_s(t,{\bf x}) := i_s(t,{\bf x}) \, {\bf 1}_{\omega}({\bf x}) \quad \text{and} \quad q_c(t,{\bf x}) := i_c(t,{\bf x}) \, {\bf 1}_{\omega}({\bf x}), 
\eeq
where ${\bf 1}_{\omega}$ is the indicator function of the set $\omega$, and $i_s(t,{\bf x})$ and $i_c(t,{\bf x})$ are the rates of inflow of oxygen and chemotherapeutic agent through intra-tumoural blood vessels at position ${\bf x} \in \omega$ and time $t$.

\section{Analysis of evolutionary dynamics}
\label{sec:analysis}
In order to obtain a comprehensive analytical description of the evolutionary dynamics of tumour cells, in this section we focus on a scenario where the concentrations of oxygen and chemotherapeutic agent are given and stationary, \emph{i.e.} when, instead of being solutions of~\eqref{eq:s} and~\eqref{eq:c}, the functions $s(t,{\bf x})$ and $c(t,{\bf x})$ are given and satisfy the following assumptions
\beq
\label{assbars}
s(t,{\bf x}) \equiv S({\bf x}) \quad \text{and} \quad c(t,{\bf x}) \equiv C({\bf x}),
\eeq
with 
\begin{equation}
S \in {\rm C}(\overline{\Omega}) \; \text{ with } \; S : \overline{\Omega} \to \mathbb{R}_{\geq0} \quad \text{and} \quad C \in {\rm C}(\overline{\Omega}) \; \text{ with } \; C : \overline{\Omega} \to \mathbb{R}_{\geq0}.
\label{assbars2}
\end{equation}
Under assumptions \eqref{assbars} and \eqref{assbars2}, we introduce the abridged notation
$$
a \equiv a(S({\bf x}),C({\bf x})), \quad b \equiv b(S({\bf x}),C({\bf x})), \quad h \equiv h(S({\bf x}),C({\bf x})).
$$
In this scenario, we construct explicit solutions of~\eqref{eq:n} ({\it cf.} Proposition~\ref{Prop1}) and we study the asymptotic behaviour of such solutions for $t \to \infty$ ({\it cf.} Theorem~\ref{Theo1}).
In agreement with much of the previous work on the mathematical analysis of the evolutionary dynamics of continuously-structured populations~\cite{perthame2006transport,rice2004evolutionary}, we focus on the case where at time $t=0$ the local phenotypic distribution of tumour cells is of the following Gaussian form 
\begin{equation}
\label{eS1S0ic}
n(0,{\bf x},y) =\rho_0({\bf x}) \sqrt{\frac{v_0({\bf x})}{2 \pi}} \, \exp\left[-\frac{v_0({\bf x})}{2} \left(y - \mu_0({\bf x}) \right)^2\right], \quad \forall \ {\bf x} \in \overline{\Omega}
\end{equation} 
where $v_0({\bf x}) := 1/\sigma^{2}_0({\bf x})$ and
\begin{equation}
\label{eS1S0icass}
\rho_0 \in {\rm C}(\overline{\Omega}) \; \text{ with } \; \rho_0 : \overline{\Omega} \to \mathbb{R}_{>0}, \quad \sigma^{2}_0 \in {\rm C}(\overline{\Omega}) \; \text{ with } \; \sigma^{2}_0 : \overline{\Omega} \to \mathbb{R}_{>0}, \quad \mu_0 \in {\rm C}(\overline{\Omega}) \; \text{ with } \; \mu_0 : \overline{\Omega} \to \mathbb{R}.
\end{equation} 

\begin{proposition}
\label{Prop1}
Let assumptions~\eqref{def:R}, \eqref{pn2ori}-\eqref{eq:h}, \eqref{assbars} and \eqref{assbars2} hold. Then, \eqref{eq:n} subject to \eqref{eS1S0ic} and~\eqref{eS1S0icass} admits the exact solution
\begin{equation}\label{gaussian}
n(t,{\bf x},y) =\rho(t,{\bf x}) \sqrt{\frac{v(t,{\bf x})}{2 \pi}} \, \exp\left[-\frac{v(t,{\bf x})}{2} \left(y - \mu(t,{\bf x}) \right)^2\right], \quad \forall \ {\bf x} \in \overline{\Omega},
\end{equation}
with $\rho(t,{\bf x})$, $\mu(t,{\bf x})$ and $v(t,{\bf x}) := 1/\sigma^{2}(t,{\bf x})$ being solutions of the Cauchy problem
\begin{equation}
\label{eq:rhomuv}
\left\{
\begin{array}{ll}
\displaystyle{\partial_t v = 2 \left(b - \beta v^2\right), \quad v \equiv v(t,{\bf x})},
\\\\
\displaystyle{\partial_t \mu = \frac{2 \ b}{v} \left(h - \mu\right), \quad \mu \equiv \mu(t,{\bf x})}, 
\\\\
\displaystyle{\partial_t \rho = \left[\left(a - \frac{b}{v} - b \left(\mu - h \right)^2 \right) - \zeta \rho \right]\rho, \quad \rho \equiv \rho(t,{\bf x})},
\\\\
v(0,{\bf x}) = 1/\sigma^{2}_0({\bf x}), \quad \mu(0,{\bf x}) = \mu_0({\bf x}), \quad \rho(0,{\bf x}) = \rho_0({\bf x}), 
\end{array}
\right.
\qquad 
(t,{\bf x}) \in (0,\infty) \times \overline{\Omega}.
\end{equation} 
\end{proposition}

\begin{theorem} \label{Theo1}
Let assumptions~\eqref{def:R}, \eqref{pn2ori}-\eqref{eq:h}, \eqref{assbars} and \eqref{assbars2} hold. Then, the solution of~\eqref{eq:n} subject to \eqref{eS1S0ic} and~\eqref{eS1S0icass} is such that 
\beq
\label{Th1i}
\rho(t,\cdot) \longrightarrow \rho_{\infty}(S,C), \quad \mu(t,\cdot) \longrightarrow \mu_{\infty}(S,C), \quad \sigma^2(t,\cdot) \longrightarrow \sigma^2_{\infty}(S,C) \quad \text{as } t \to \infty,
\eeq
with
\beq
\label{Th1ii}
\rho_{\infty}(S,C) = \max\Bigg(0,\frac{a(S,C)-\sqrt{\beta\,b(S,C)}}{\zeta }\Bigg), \quad \mu_{\infty}(S,C) = h(S,C), \quad \sigma^2_{\infty}(S,C) =\sqrt{\frac{\beta}{b(S,C)}}.
\eeq
\end{theorem}

\begin{figure*}[h!]
  \includegraphics[width=\textwidth]{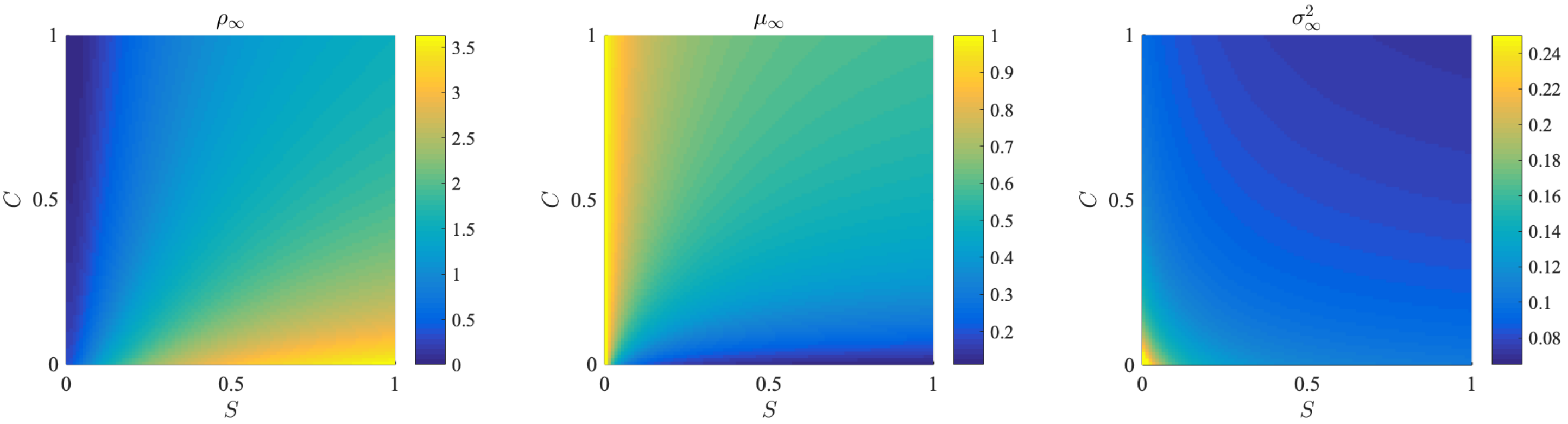}
\caption{{\bf Plots of $\rho_{\infty}(S,C)$, $\mu_{\infty}(S,C)$ and $\sigma^2_{\infty}(S,C)$.} Plots of the equilibrium cell density $\rho_{\infty}$, the equilibrium local mean phenotypic state $\mu_{\infty}$ and the related variance $\sigma^2_{\infty}$ defined via \eqref{Th1ii} as functions of the stationary  concentrations of oxygen $S$ and chemotherapeutic agent $C$. The plots refer to the parameter values listed in Table~\ref{Tab1}. The cell density is in units of $10^8$ and the concentrations of oxygen and chemotherapeutic agent are scaled by the reference values $S_0$ and $C_0$ given in Table~\ref{Tab1}, respectively.}
\label{fig:0}       
\end{figure*}

The proofs of Proposition~\ref{Prop1} and Theorem~\ref{Theo1} are reported in Appendix~\ref{Prop1:proof} and Appendix~\ref{Theo1:proof}, respectively. These results provide a mathematical formalisation of the idea that, when the concentrations of oxygen and chemotherapeutic agent are given and stationary (\emph{i.e.} $s(t,{\bf x})\equiv S({\bf{x}})$ and $c(t,{\bf x})\equiv C({\bf{x}})$), the tumour cell density $\rho(t,{\bf x})$, the local mean phenotypic state $\mu(t,{\bf x})$ and the related variance $\sigma^2(t,{\bf x})$ converge to some equilibrium values which are determined by the local concentrations of oxygen and chemotherapeutic agent. This is illustrated by the heat maps in Figure~\ref{fig:0}, which show how, for the biologically consistent parameter values listed in Table~\ref{Tab1}, the values of $\rho_{\infty}$, $\mu_{\infty}$ and $\sigma^2_{\infty}$ defined via \eqref{Th1ii} vary as functions of $S$ and $C$. Notice that the parameter values in Table~\ref{Tab1} are such that $\rho_{\infty}>0$.

These results demonstrate that spatial variations of the oxygen concentration determine spatial variations of the tumour cell density, of the local mean phenotypic state and the related variance. Specifically, under the parameter values listed in Table~\ref{Tab1}, the tumour cell density $\rho_{\infty}$ is an increasing function of the oxygen concentration. Moreover, the local mean phenotypic state $\mu_{\infty}$ coincides with the fittest phenotypic state $h$, which decreases from values close to $y=1$ (\emph{i.e.} the state corresponding to the phenotypic variant with the highest rate of cellular division via anaerobic energy pathways) to values close to $y=0$ (\emph{i.e.} the state corresponding to the phenotypic variant with the highest rate of cellular division via aerobic energy pathways) for increasing values of the oxygen concentration. This suggests that aerobic phenotypic variants are to be expected to colonise oxygenated regions of the tumour, while anaerobic phenotypic variants are likely to populate poorly-oxygenated regions. Finally, the local phenotypic variance $\sigma^2_{\infty}$ is a decreasing function of the oxygen concentration, which supports the idea that higher levels of phenotypic variability may occur in hypoxic regions of the tumour. 

On the other hand, larger values of the concentration of chemotherapeutic agent bring about smaller values of the tumour cell density $\rho_{\infty}$, a shift of the local mean phenotypic state $\mu_{\infty}$ (\emph{i.e.} the fittest phenotypic state $h$) from values closer to $y=0$ to values closer to $y=1$ (\emph{i.e.} the state corresponding to the anaerobic phenotypic variant with the highest level of resistance to chemotherapy), and smaller values of the local phenotypic variance $\sigma^2_{\infty}$. This indicates that the selective pressure exerted by the chemotherapeutic agent causes a population bottleneck in tumour cells leading to a reduction in cell density coming along with the selection of more chemoresistant phenotypic variants and lower levels of phenotypic variability.
\begin{remark}
\label{rem1}
Under the assumptions of Theorem~\ref{Theo1}, in the case where~\eqref{eq:s} and~\eqref{eq:c} subject to~\eqref{eq:bcs} and coupled with~\eqref{eq:n} admit classical solutions $s(t,{\bf x})$ and $c(t,{\bf x})$ that converge to some limits $s_{\infty}({\bf x})$ and $c_{\infty}({\bf x})$ as $t \to \infty$, we expect the long-time asymptotic limit of the local phenotypic distribution of tumour cells $n(t,{\bf x},y)$ to be of the Gaussian form
\beq
\label{Th1ir1}
n_{\infty}({\bf x},y) = \frac{\rho_{\infty}({\bf x})}{\sqrt{2\pi \, \sigma^{2}_{\infty}({\bf x})}} \,\exp \Bigg[ -\frac{1}{2 \, \sigma^2_{\infty}({\bf x})}\big(y-\mu_{\infty}({\bf x})\big)^2\Bigg]
\eeq
where
\beq
\label{Th1iir1}
\rho_{\infty}({\bf x}) = \max\Bigg(0,\frac{a(s_{\infty}({\bf x}),c_{\infty}({\bf x})) - \sqrt{\beta\,b(s_{\infty}({\bf x}),c_{\infty}({\bf x}))}}{\zeta }\Bigg), 
\eeq
\beq
\label{Th1iiir1}
\mu_{\infty}({\bf x}) = h(s_{\infty}({\bf x}),c_{\infty}({\bf x})) \quad \text{and} \quad \sigma^{2}_{\infty}({\bf x}) =\sqrt{\frac{\beta}{b(s_{\infty}({\bf x}),c_{\infty}({\bf x}))}}.
\eeq
\end{remark}

\section{Numerical simulations}
\label{sec:numsol}
We complement the analytical results of evolutionary dynamics presented in Section~\ref{sec:analysis} with numerical solutions of the model equations. In Section~\ref{sec:setup}, we describe the set-up of numerical simulations and the methods employed to construct numerical solutions. In Section~\ref{sec:numsol:constant}, we consider the case of a one-dimensional spatial domain whereby the concentrations of oxygen and chemotherapeutic agent are stationary. In Section~\ref{sec:numsol:evolving}, we focus on the case of a two-dimensional spatial domain and let the dynamics of oxygen and chemotherapeutic agent be governed by~\eqref{eq:s} and \eqref{eq:c}. All simulations are carried out using the parameter values listed in Table~\ref{Tab1}, which are chosen to be consistent with the existing literature. 

\subsection{\bf Set-up of numerical simulations and numerical methods}
\label{sec:setup}
\begin{figure*}[h!]
\centering
  \includegraphics[width=\textwidth]{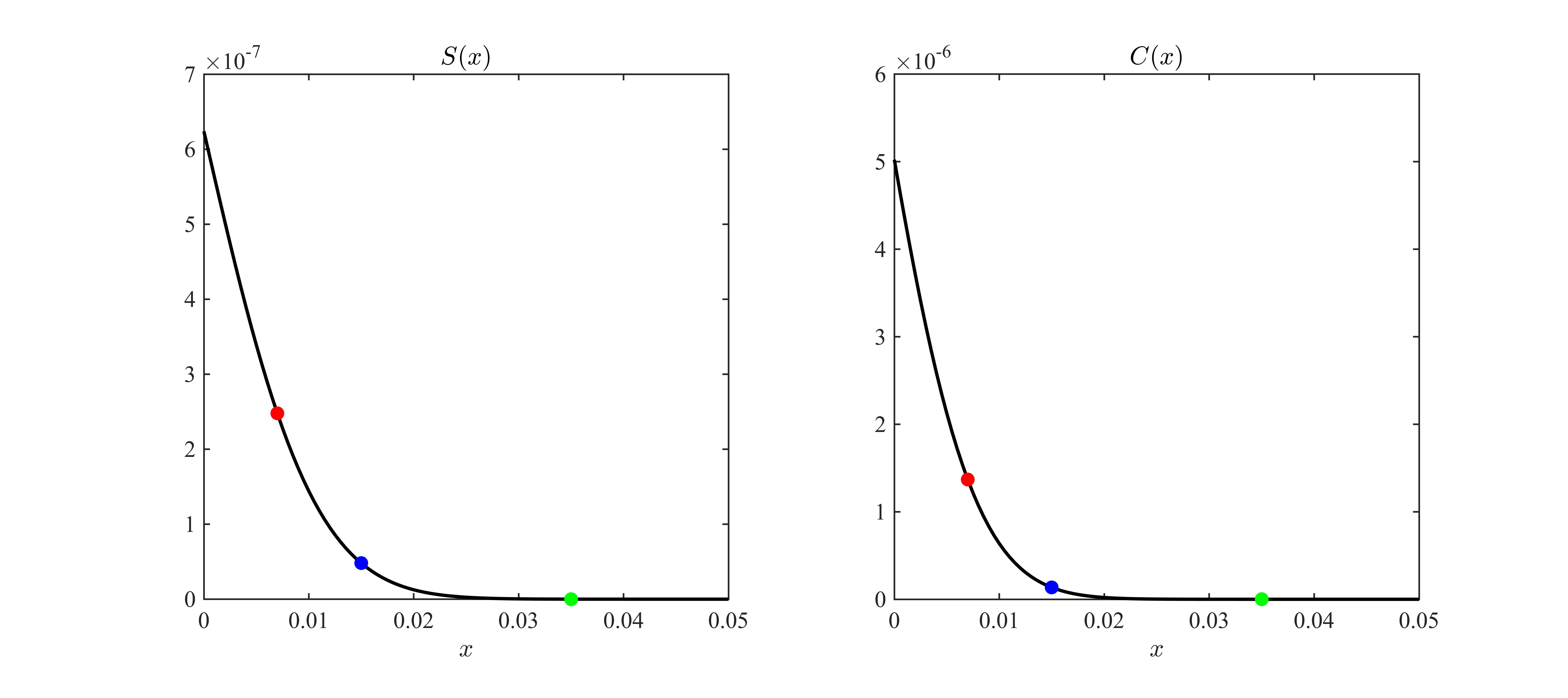}
\caption{{\bf Stationary concentrations of oxygen and chemotherapeutic agent considered in Section~\ref{sec:numsol:constant}.} Plots of the oxygen concentration $S(x)$ and the concentration of chemotherapeutic agent $C(x)$ used to obtain the numerical results of Figure~\ref{fig:2} and Figure~\ref{fig:3}. 
The coloured dots highlight the values of $S(x)$ and $C(x)$ corresponding to the lines of the same colours in Figure~\ref{fig:2} and Figure~\ref{fig:3} -- \emph{i.e.} $S(x)$ and $C(x)$ at $x=0.007$ (red), $x=0.015$ (blue) and $x=0.035$ (green). The space variable $x$ is in units of $cm$, while both $S(x)$ and $C(x)$ are in units of $g\,cm^{-3}$. The oxygen concentration $S(x)$ is defined in such a way as to match the experimental pO$_2$ profile presented in~\cite[Fig. 3]{helmlinger1997interstitial}. The conversion from $mmHg$ of pO$_2$ to $g\,cm^{-3}$ of oxygen concentration was performed using the conversion factor $1\,mmHg=4.6\times 10^{-8}\,g\,cm^{-3}$, which was estimated using the ideal gas law. The concentration of chemotherapeutic agent $C(x)$ is defined in such a way as to have a behaviour which is qualitatively similar to that of $S(x)$ and the value of $C(0)$ is chosen in agreement with experimental data presented in~\cite{helmlinger1997interstitial}. 
 }
\label{fig:1}       
\end{figure*}

\paragraph{\bf Set-up of numerical simulations of Section~\ref{sec:numsol:constant}.} For the numerical simulations we report on in Section~\ref{sec:numsol:constant}, we define $\Omega:=(0,0.05)$ and assume that increasing values of ${\bf x} \equiv x$ correspond to increasing values of the distance from a blood vessel located in $x=0$. Under the parameter values listed in Table~\ref{Tab1}, the values of $x$ are in units of $cm$. Under assumptions~\eqref{assbars} and~\eqref{assbars2}, we define $S(x)$ and $C(x)$ as shown by the plots in Figure~\ref{fig:1}. Here, the stationary oxygen concentration $S(x)$ is defined in such a way as to match the experimental oxygen distribution presented in~\cite[Fig. 3]{helmlinger1997interstitial}. Furthermore, the stationary concentration of chemotherapeutic agent $C(x)$ is defined in such a way as to have a behaviour qualitatively similar to that of $S(x)$ and the value of $C(0)$ is chosen in agreement with experimental data presented in~\cite{helmlinger1997interstitial}.
\\\\
We complement~\eqref{eq:n} with initial condition~\eqref{eS1S0ic} and assume
\begin{equation}\label{num:ic1}
\sigma^2(0,{\bf x}) \equiv \sigma^2_0 = 1 \,, \quad \mu(0,{\bf x}) \equiv \mu_0 = 0.5 \quad \text{and} \quad \rho(0,{\bf x}) \equiv \rho_0 \approx 10^8. 
\end{equation}
Assumptions~\eqref{num:ic1} correspond to a biological scenario whereby at the initial time $t=0$ tumour cells are uniformly distributed across the spatial domain $\Omega$ and are mainly found in the phenotypic state $y=0.5$. 

\paragraph{\bf Set-up of numerical simulations of Section~\ref{sec:numsol:evolving}.} For the numerical simulations we report on in Section~\ref{sec:numsol:evolving}, we define $\Omega:=(0,0.5)\times(0,0.5)$ in order to model the cross-section of a vascularised tumour tissue. Under the parameter values listed in Table~\ref{Tab1}, the values of ${\bf x} \in \Omega$ are in units of $cm$. We let the dynamics of oxygen and chemotherapeutic agent be governed by~\eqref{eq:s}-\eqref{eq:bcs}. Moreover, we assume the rate of inflow of oxygen and chemotherapeutic agent through intra-tumoural blood vessels to be constant in time and the same for all vessels, \emph{i.e.} we define the functions $i_s(t,{\bf x})$ and $i_c(t,{\bf x})$ in~\eqref{q} as
\beq
\label{constS}
i_s(t,{\bf x}) \equiv I_s  \quad \text{and} \quad i_c(t,{\bf x}) \equiv I_c,
\eeq
with the values of $I_s$ and $I_c$ being those given in Table~\ref{Tab1}.
\\\\
We complement~\eqref{eq:n} with the initial condition defined via~\eqref{eS1S0ic} and~\eqref{num:ic1}, while \eqref{eq:s} and \eqref{eq:c} are complemented with the following initial conditions
\begin{equation}\label{num:ic3}
s(0,{\bf x}) = S_0\, {\bf 1}_{\omega}({\bf x}) \quad \text{and} \quad  c(0,{\bf x}) = C_0\, {\bf 1}_{\omega}({\bf x}),
\end{equation}
with the values of $S_0$ and $C_0$ being those given in Table~\ref{Tab1}. These initial conditions correspond to a biological scenario whereby at the initial time $t=0$ tumour cells are uniformly distributed across the spatial domain $\Omega$ and are mainly found in the phenotypic state $y=0.5$, while the oxygen and the chemotherapeutic agent are concentrated in correspondence of the blood vessels. 

\begin{table*}[tbhp]
	{\footnotesize
		\centering
		\caption{Parameter values used in numerical simulations}
		\hspace{-0.7cm}\begin{tabular}{cllc}
			{\bf Parameter} & {\bf Biological meaning} & {\bf Value} & {\bf Reference} \\
			$\alpha_c$ & Michaelis-Menten constant of chemotherapeutic agent & $2 \times 10^{-6}$ $g \, cm^{-3}$ & \cite{norris2006modelling} \\ 
			$\alpha_s$ & Michaelis-Menten constant of oxygen & $1.5 \times 10^{-7}$ $g \, cm^{-3}$ & \cite{casciari1992variations} \\ 
			$\beta$ & Rate of spontaneous epimutation & $10^{-6}$ $s^{-1}$ & \cite{chisholm2015emergence} \\
			$D_c$ & Diffusivity of chemotherapeutic agent & $4.4 \times 10^{-6}$ $cm^2 \, s^{-1}$ & \cite{powathil2012modelling} \\
			$D_s$ & Diffusivity of oxygen & $2 \times 10^{-5}$ $cm^2 \, s^{-1}$ & \cite{hlatky1985two} \\
			$\gamma_c$ & Maximal reduction of bkg fitness of aerobic phenotype due to chemotherapy & $1.8 \times 10^{-4}$ $s^{-1}$ & \cite{ward1997mathematical} \\ 
			$\gamma_s$ & Maximal bkg fitness of aerobic phenotype & $1 \times 10^{-4}$ $s^{-1}$ & \cite{ward1997mathematical} \\ 
			$\zeta$ & Rate of cell death due to competition for space  & $2 \times 10^{-13}$ $cm^{3} \, s^{-1} \, cells^{-1}$ & \cite{li1982glucose}  \\ 
			$\eta_c$ & Conversion factor for cell consumption of chemotherapeutic agent & $4 \times 10^{-11}$ $g \, cells^{-1}$ & \cite{norris2006modelling} \\
		        $\eta_s$ & Conversion factor for cell consumption of oxygen & $2 \times 10^{-11}$ $g \, cells^{-1}$ & \cite{casciari1992variations} \\
			$\lambda_c$ & Rate of natural decay of chemotherapeutic agent & $2.3 \times 10^{-4}$ $s^{-1}$& \cite{powathil2012modelling}  \\ 
			$\lambda_s$ & Rate of natural decay of oxygen & $2.78 \times 10^{-6}$ $s^{-1}$& \cite{cumsille2015proposal}  \\ 
			$\varphi$ & Maximal bkg fitness of anaerobic phenotype & $1 \times 10^{-5}$ $s^{-1}$ & \cite{gordan2007hif} \\ 
			$I_c$ & Constant rate of inflow of chemotherapeutic agent through blood vessels & $2.5 \times 10^{-6}$ $g \, cm^{-3} \, s^{-1}$ & \cite{norris2006modelling}  \\
			$I_s$ & Constant rate of inflow of oxygen through blood vessels & $6.3996 \times 10^{-7}$ $g \, cm^{-3} \, s^{-1}$ & \cite{kumosa2014permeability}  \\ 
		         $C_0$ & Reference value for the concentration of chemotherapeutic agent & $2.5 \times 10^{-6}$ $g \, cm^{-3}$ & \cite{norris2006modelling}  \\
			$S_0$ & Reference value for the concentration of oxygen & $6.3996 \times 10^{-7}$ $g \, cm^{-3}$ & \cite{kumosa2014permeability}   
		\end{tabular}
		\label{Tab1}
	}
\end{table*}

\paragraph{\bf Numerical methods.} Numerical solutions are constructed using a uniform discretisation of the interval $[0,0.05]$ or the square $[0,0.5]\times[0,0.5]$ as the computational domain of the independent variable ${\bf x}$. Moreover, a uniform discretisation of the set $[-7,7]$ is used as the computational domain of the independent variable $y$. We consider $t \in [0,{\rm T}]$, with ${\rm T}>0$ being the final time of simulations. The final time ${\rm T}$ is chosen sufficiently large so as to ensure that the solutions are at numerical equilibrium at the end of simulations. The exact values of ${\rm T}$ are reported in the captions of Figures~\ref{fig:2}-\ref{fig:7}. We discretise the interval $[0,{\rm T}]$ with a uniform step. The method for solving numerically~\eqref{eq:n} subject to the zero-flux boundary conditions
$$
\partial_y n(\cdot, \cdot, -7) = 0 \quad  \text{and} \quad \partial_y n(\cdot, \cdot, 7) = 0 
$$
is based on an explicit finite difference scheme in which a three-point stencil is used to approximate the diffusion term in $y$ and an explicit finite difference scheme is used for the non-local reaction term. Furthermore, the method for solving numerically~\eqref{eq:s} and \eqref{eq:c} subject to the zero-flux boundary conditions~\eqref{eq:bcs} is based on an explicit finite difference scheme whereby a five-point stencil is used to approximate the diffusion terms and an explicit finite difference scheme is used for the other terms. Finally, numerical solutions to the Cauchy problem~\eqref{eq:rhomuv} are constructed using the explicit Euler method. All numerical computations are performed in {\sc Matlab}.

\subsection{\bf One-dimensional numerical results under stationary concentrations of oxygen and chemotherapeutic agent}
\label{sec:numsol:constant}
\begin{figure*}[h!]
 \includegraphics[width=\textwidth]{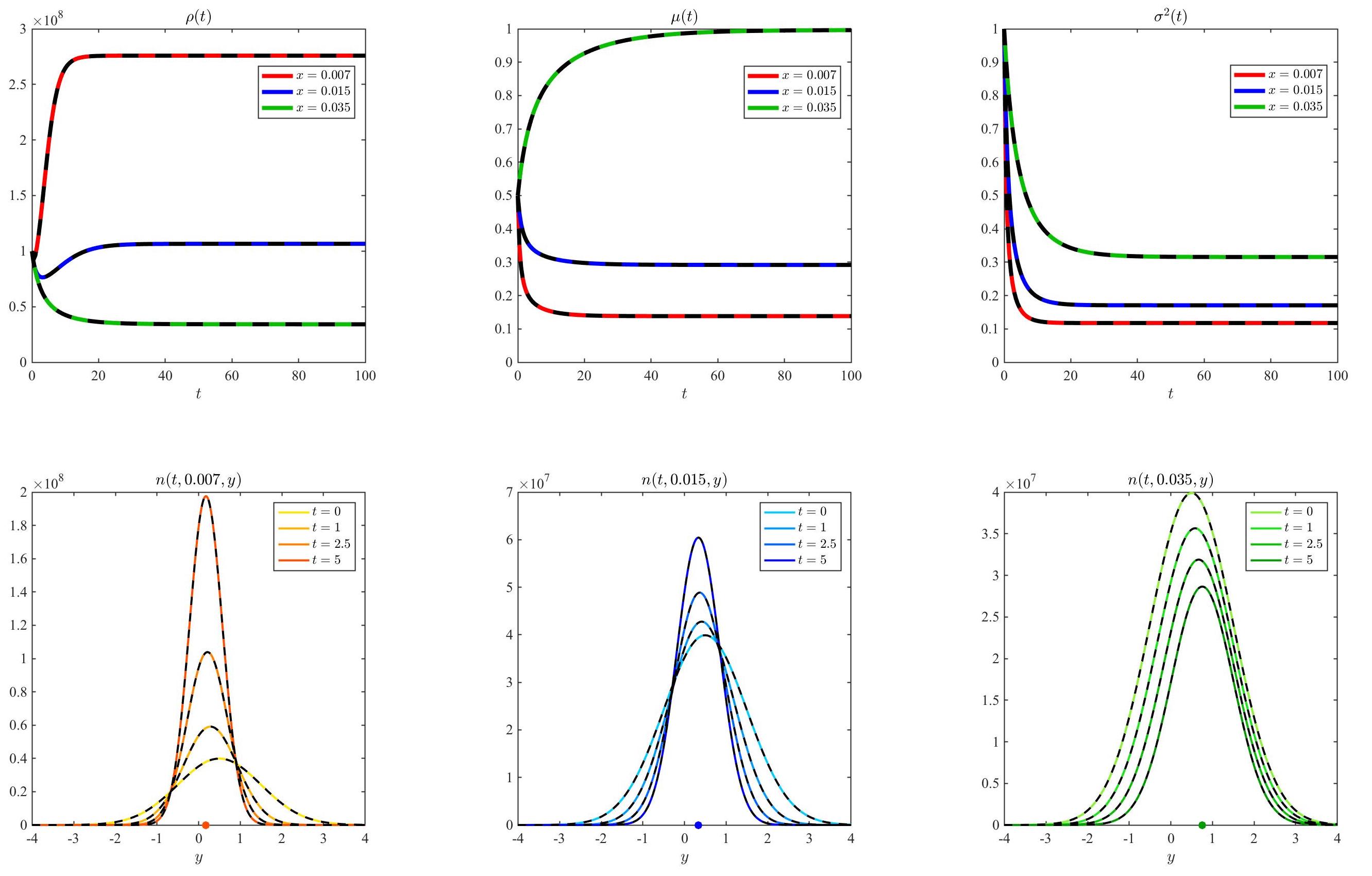}
\caption{\textbf{One-dimensional numerical results under stationary concentration of oxygen and in the absence of chemotherapeutic agent.} {\it Top row.} Plots of the cell density $\rho(t,x)$ (left panel), the local mean phenotypic state $\mu(t,x)$ (central panel) and the related variance $\sigma^2(t,x)$ (right panel) at $x=0.007$ (red, solid lines), $x=0.015$ (blue, solid lines) and $x=0.035$ (green, solid lines) obtained by solving numerically~\eqref{eq:n} subject to the initial condition defined via~\eqref{eS1S0ic} and~\eqref{num:ic1}, under the stationary concentration of oxygen $s(t,x) \equiv S(x)$ displayed in Figure~\ref{fig:1} and the stationary concentration of chemotherapeutic agent $c(t,x) \equiv 0$ (\emph{i.e.} in the absence of chemotherapeutic agent). The black, dashed lines highlight the corresponding quantities obtained by solving numerically the Cauchy problem~\eqref{eq:rhomuv} complemented with~\eqref{num:ic1}.  {\it Bottom row.} Plots of the local cell phenotypic distribution $n(t,x,y)$ obtained by solving numerically~\eqref{eq:n} subject to the initial condition defined via~\eqref{eS1S0ic} and~\eqref{num:ic1}, under the stationary concentration of oxygen $s(t,x) \equiv S(x)$ displayed in Figure~\ref{fig:1} and the stationary concentration of chemotherapeutic agent $c(t,x) \equiv 0$ (\emph{i.e.} in the absence of chemotherapeutic agent), at $x=0.007$ (left panel), $x=0.015$ (central panel) and $x=0.035$ (right panel). Different solid, coloured lines correspond to different time instants $t$ and the dashed lines highlight the exact solution~\eqref{gaussian} with $\sigma^2(t,x)$, $\mu(t,x)$ and $\rho(t,x)$ given by numerical solutions of the Cauchy problem~\eqref{eq:rhomuv} complemented with~\eqref{num:ic1}. The bullets on the axis of abscissas highlight the value of the mean phenotypic state $\mu(t,x)$ at $t=5$. The time variable $t$ is in units of $10^4 \, s$, the space variable $x$ is in units of $cm$ and the parameters values used are those listed in Table~\ref{Tab1}.}
\label{fig:2}       
\end{figure*}

\begin{figure*}[h!]
 \includegraphics[width=\textwidth]{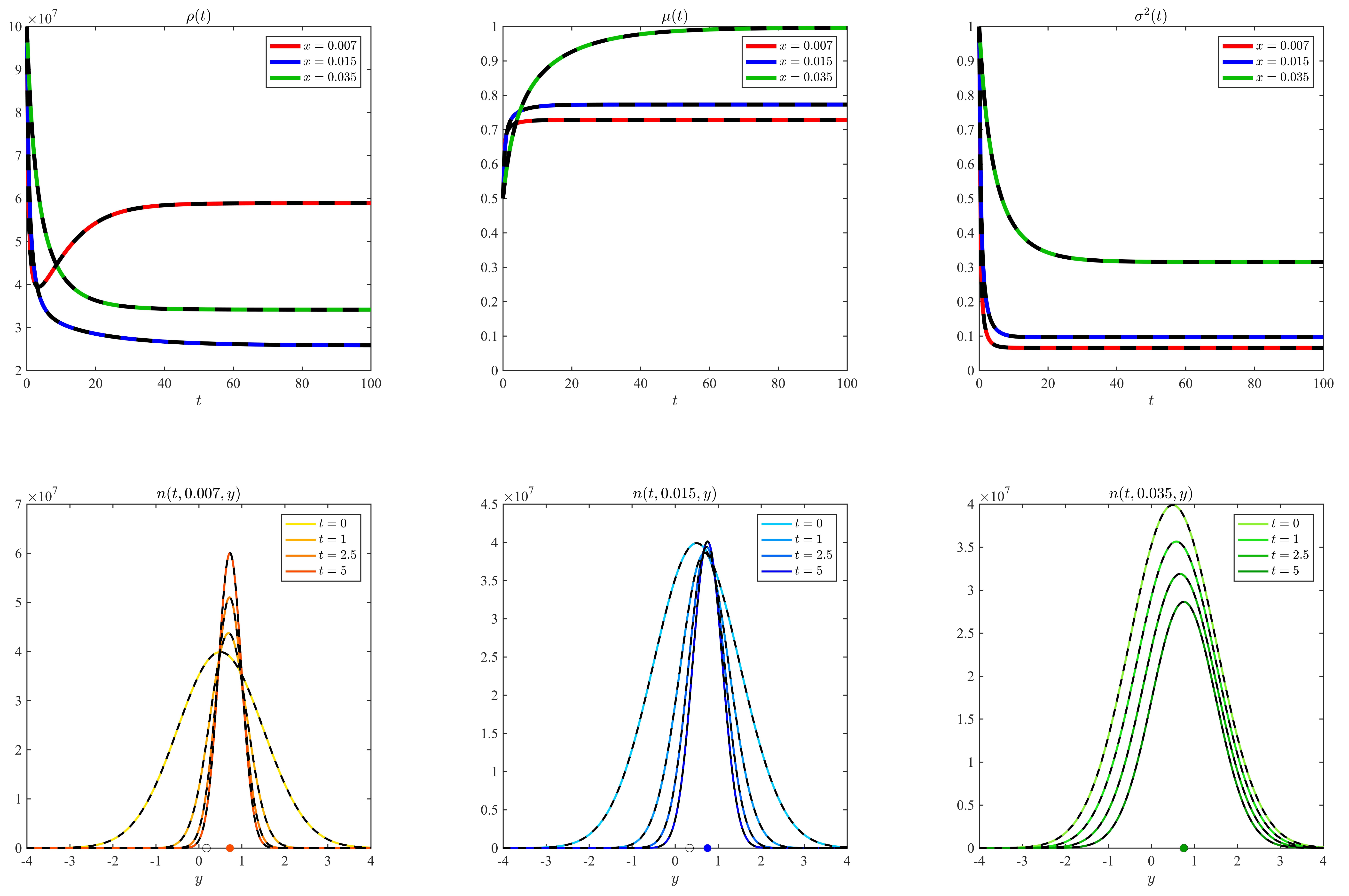}
\caption{\textbf{One-dimensional numerical results under stationary concentrations of oxygen and chemotherapeutic agent.} {\it Top row.} Plots of the cell density $\rho(t,x)$ (left panel), the local mean phenotypic state $\mu(t,x)$ (central panel) and the related variance $\sigma^2(t,x)$ (right panel) at $x=0.007$ (red, solid lines), $x=0.015$ (blue, solid lines) and $x=0.035$ (green, solid lines) obtained by solving numerically~\eqref{eq:n} subject to the initial condition defined via~\eqref{eS1S0ic} and~\eqref{num:ic1}, and under the stationary concentrations of oxygen $s(t,x) \equiv S(x)$ and chemotherapeutic agent $c(t,x) \equiv C(x)$ displayed in Figure~\ref{fig:1}. The black, dashed lines highlight the corresponding quantities obtained by solving numerically the Cauchy problem~\eqref{eq:rhomuv} complemented with~\eqref{num:ic1}.  {\it Bottom row.} Plots of the local cell phenotypic distribution $n(t,x,y)$ obtained by solving numerically~\eqref{eq:n} subject to the initial condition defined via~\eqref{eS1S0ic} and~\eqref{num:ic1}, and under the stationary concentrations of oxygen $S(x)$ and chemotherapeutic agent $C(x)$ displayed in Figure~\ref{fig:1}, at $x=0.007$ (left panel), $x=0.015$ (central panel) and $x=0.035$ (right panel). Different solid, coloured lines correspond to different time instants $t$ and the dashed lines highlight the exact solution~\eqref{gaussian} with $\sigma^2(t,x)$, $\mu(t,x)$ and $\rho(t,x)$ given by numerical solutions of the Cauchy problem~\eqref{eq:rhomuv} complemented with~\eqref{num:ic1}. The filled bullets on the axis of abscissas highlight the value of the mean phenotypic state $\mu(t,x)$ at $t=5$, while the empty bullets highlight the corresponding values obtained in the case where $c(t,x) \equiv 0$ (\emph{i.e.} in the absence of chemotherapeutic agent). The time variable $t$ is in units of $10^4 \, s$, the space variable $x$ is in units of $cm$ and the parameters values used are those listed in Table~\ref{Tab1}.}
\label{fig:3}       
\end{figure*}

The sample of numerical results presented in Figure~\ref{fig:2} refer to the case where the oxygen concentration $s(t,x) \equiv S(x)$ and the concentration of cytotoxic agent $c(t,x) \equiv 0$, while the results presented in Figure~\ref{fig:3} refer to the case where $s(t,x) \equiv S(x)$ and $c(t,x) \equiv C(x)$, with $S(x)$ and $C(x)$ being defined as illustrated by the plots in Figure~\ref{fig:1}. 

\paragraph{\bf Agreement between analytical and numerical results.} In agreement with the results established by Proposition~\ref{Prop1}, the numerical results displayed in the top rows of Figure~\ref{fig:2} and Figure~\ref{fig:3} 
show that there is a perfect match between the cell density $\rho(t,x)$, the local mean phenotypic state $\mu(t,x)$ and the related variance $\sigma^2(t,x)$ computed via numerical integration of the local cell phenotypic distribution $n(t,x,y)$, which is obtained by solving numerically~\eqref{eq:n} subject to the initial condition defined via~\eqref{eS1S0ic} and~\eqref{num:ic1}, and the corresponding quantities obtained by solving numerically the Cauchy problem~\eqref{eq:rhomuv} complemented with~\eqref{num:ic1}. Similarly, the sample of numerical results presented in the bottom rows of Figure~\ref{fig:2} and Figure~\ref{fig:3} show that the local cell phenotypic distribution $n(t,x,y)$ matches the exact local cell phenotypic distribution~\eqref{gaussian}. Moreover, in accordance with the asymptotic results established by Theorem~\ref{Theo1}, the cell density, the local mean phenotypic state and the related variance converge, respectively, to the equilibrium values $\rho_{\infty}({\bf x})$, $\mu_{\infty}({\bf x})$ and $\sigma^{2}_{\infty}({\bf x})$ given by~\eqref{Th1ii}.

\paragraph{\bf Tumour cell dynamics in the absence of chemotherapeutic agent.} The numerical results of Figure~\ref{fig:2} show that, in the absence of chemotherapeutic agent, since the stationary oxygen concentration $S(x)$ decreases monotonically with the distance from the blood vessel located at $x=0$ ({\it vid.} Figure~\ref{fig:1}), the cell density $\rho(t,x)$ at equilibrium is maximal in the vicinity of the blood vessel ({\it cf.} red line), where the oxygen concentration is higher, and decreases monotonically as the distance from the vessel increases ({\it cf.} blue and green lines). Accordingly, the local mean phenotypic state at equilibrium increases from values closer to $y=0$ (\emph{i.e.} the state corresponding to the phenotypic variant with the highest rate of cellular division via aerobic energy pathways) to values closer to $y=1$ (\emph{i.e.} the state corresponding to the phenotypic variant with the highest rate of cellular division via anaerobic energy pathways) moving away from the blood vessel. Moreover, the local phenotypic variance $\sigma^2(t,x)$ at equilibrium is a monotonically increasing function of the distance from the blood vessel (\emph{i.e.} local phenotypic variability increases with the distance from the blood vessel). 

\paragraph{\bf Tumour cell dynamics in the presence of chemotherapeutic agent.} A comparison of the numerical results of Figure~\ref{fig:2} and Figure~\ref{fig:3} reveals that in the regions in close proximity of the blood vessel ({\it cf.} red lines), where its concentration is higher, the chemotherapeutic agent leads to the occurrence of a population bottleneck in tumour cells, which results into: a reduction of the equilibrium value of the cell density $\rho(t,x)$; a selective sweep toward more resistant phenotypic variants, as demonstrated by the fact that the equilibrium value of the local mean phenotypic state $\mu(t,x)$ shifts from values closer to $y=0$ (\emph{i.e.} the state corresponding to the phenotypic variant with the highest rate of cellular division via aerobic energy pathways) to values closer to $y=1$ (\emph{i.e.} the state corresponding to the anaerobic phenotypic variant with the highest level of resistance to chemotherapy); a reduction of the equilibrium value of the local phenotypic variance $\sigma^2(t,x)$. Moreover, moving away from the blood vessel, since its concentration decreases, the chemotherapeutic agent has a weaker impact on the dynamics of tumour cells ({\it cf.} blue lines). As a result, the evolution of tumour cells in regions distal to the blood vessel is hardly affected by the chemotherapeutic agent ({\it cf.} green lines).

\begin{figure*}[h!]
  \includegraphics[width=\textwidth]{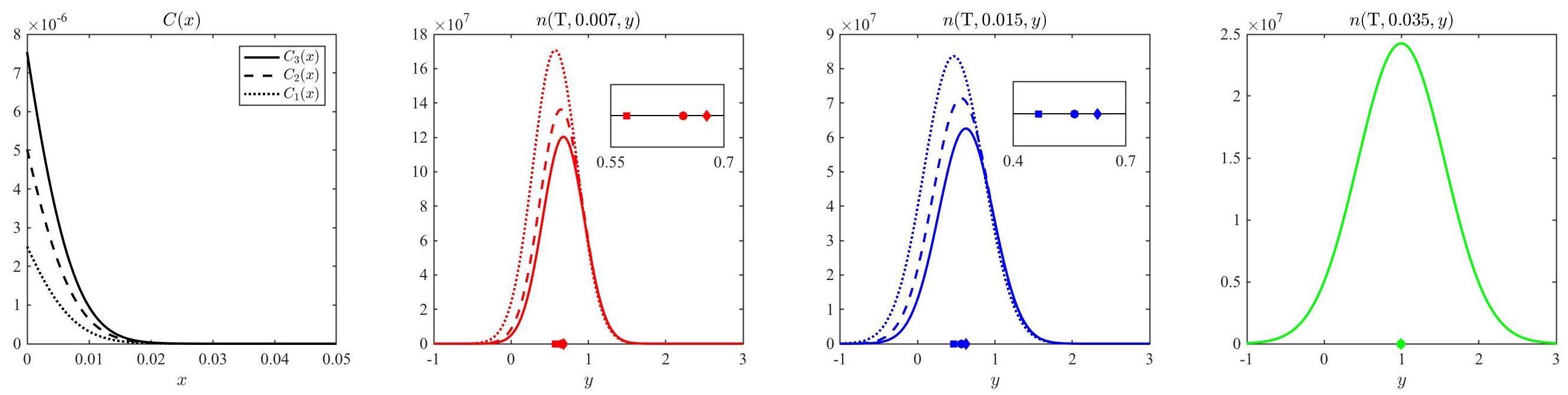}
\caption{\textbf{One-dimensional numerical results for different delivered doses of chemotherapeutic agent.} Plots of the local cell phenotypic distributions $n(${\rm T}$,x,y)$ at $x=0.007$ (second panel), $x=0.015$ (third panel) and $x=0.035$ (fourth panel) obtained by solving numerically~\eqref{eq:n} subject to the initial condition defined via~\eqref{eS1S0ic} and~\eqref{num:ic1}, under the stationary concentration of oxygen $S(x)$ displayed in Figure~\ref{fig:1} and different stationary concentrations of chemotherapeutic agent. In particular, the three stationary concentrations of chemotherapeutic agent displayed in the first panel are used, that is, $C_1(x) = 0.5\, C(x)$ (dotted line), $C_2(x) = C(x)$ (dashed line) and $C_3(x) = 1.5\, C(x)$ (solid line), where $C(x)$ is the reference concentration of chemotherapeutic agent displayed in Figure~\ref{fig:1}. In the second, third and fourth panels, the local cell phenotypic distributions at $t={\rm T}$ corresponding to $C_1$ (dotted lines), $C_2$ (dashed lines) and $C_3$ (solid lines) are displayed, and the markers on the axis of abscissas highlight the value of the mean phenotypic state $\mu({\rm T},x)$ corresponding to $C_1$ (square), $C_2$ (bullet) and $C_3$ (diamond). The insets in the second and third panel display a close-up of the axis of abscissas. The space variable $x$ is in units of $cm$, ${\rm T}= 10^6 \,s$ and the parameters values used are those listed in Table~\ref{Tab1}.}
\label{fig:3.5}       
\end{figure*}

\paragraph{\bf Tumour cell dynamics for different delivered doses of chemotherapeutic agent.}
The numerical results of Figure~\ref{fig:3.5} reproduce a realistic scenario whereby variation in the delivered dose of the chemotherapeutic agent leads to pronounced changes in the agent concentration in close proximity of the blood vessel while leaving the concentration far from the blood vessel almost unchanged (\emph{vid.} the stationary distributions of chemotherapeutic agent displayed in the first panel of Figure~\ref{fig:3.5}). These results indicate that increasing the value of the delivered dose leads to a reduction in the number of tumour cells at the cost of promoting a selective sweep toward more resistant phenotypic variants in the vicinity of the blood vessel -- \emph{i.e.} for values of $x$ sufficiently close to $0$, the area under the curve of the equilibrium local cell phenotypic distribution shrinks (\emph{vid.} the plots in the second and third panel of Figure~\ref{fig:3.5}) and the equilibrium value of the local mean phenotypic state progressively shifts from values closer to $y=0$ to values closer to $y=1$ (\emph{vid.} the insets in the second and third panel of Figure~\ref{fig:3.5}). This supports the idea that higher doses of chemotherapeutic agent removes the selective barrier limiting the growth of less proliferative and more resistant phenotypic variants in vascularised areas of the tumour.

\subsection{\bf Two-dimensional numerical results under dynamical concentrations of oxygen and chemotherapeutic agent}
\label{sec:numsol:evolving}
\begin{figure*}[h!]
\centering
  \includegraphics[width=0.95\textwidth]{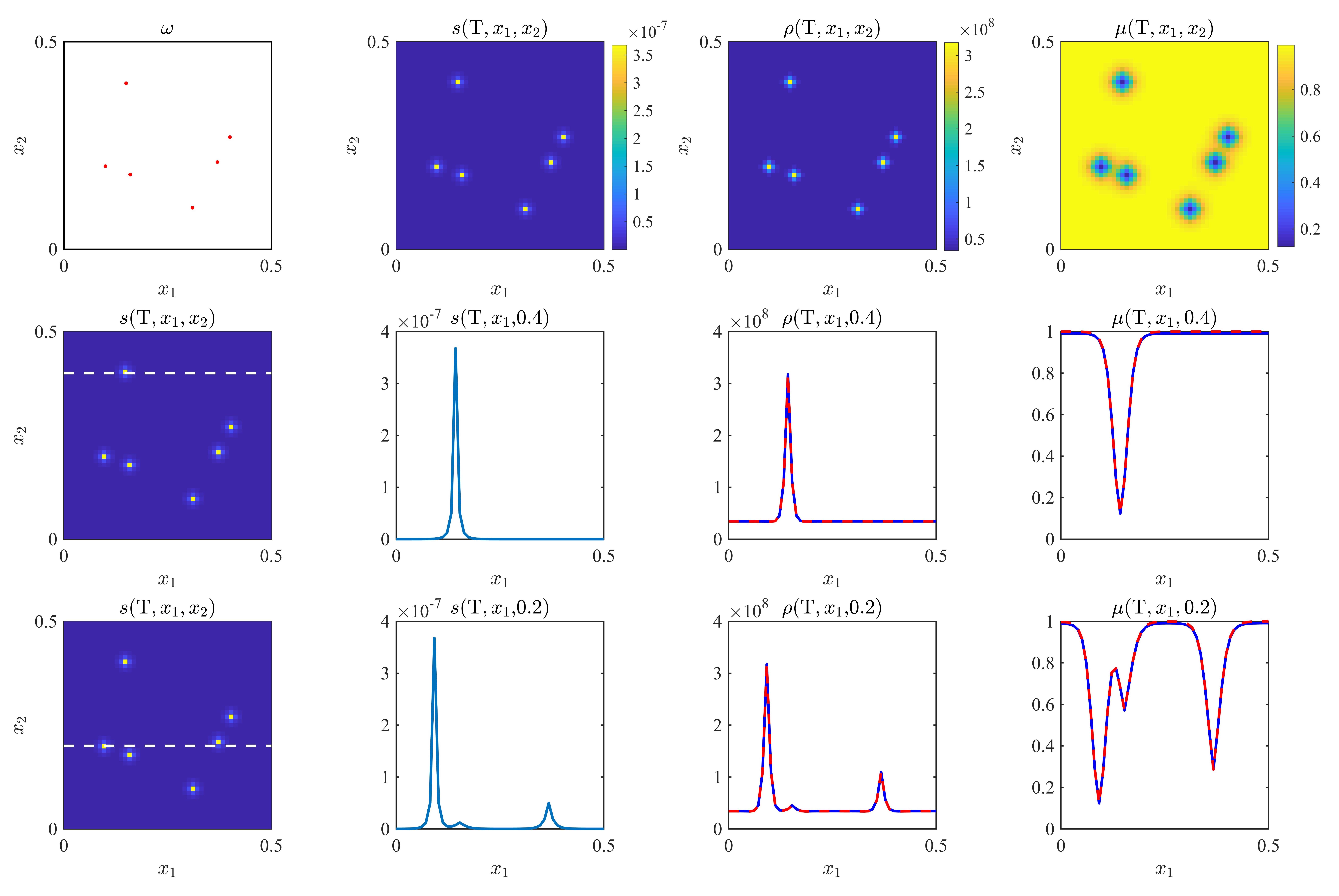}
 \caption{\textbf{Two-dimensional numerical results under dynamical concentration of oxygen and in the absence of chemotherapeutic agent.} \textit{Top row.} Plots of the oxygen concentration $s({\rm T},x_1,x_2)$ (second panel), the cell density $\rho({\rm T}, x_1,x_2)$ (third panel) and the local mean phenotypic state $\mu({\rm T},x_1,x_2)$ (fourth panel), with ${\rm T} = 5 \times 10^5 s$, obtained by solving numerically~\eqref{eq:n} and~\eqref{eq:s} imposing the initial conditions defined via~\eqref{eS1S0ic},~\eqref{num:ic1} and~\eqref{num:ic3}, the boundary condition~\eqref{eq:bcs} and assuming $c(t,x_1,x_2) \equiv 0$ (\emph{i.e.} in the absence of chemotherapeutic agent). The set $\omega$ in~\eqref{q} consists of the parts of $\Omega$ highlighted in red in the first panel. \textit{Central row.} Plots of the oxygen concentration $s({\rm T},x_1, 0.4)$ (second panel), the cell density $\rho({\rm T}, x_1, 0.4)$ (third panel, blue line) and the local mean phenotypic state $\mu({\rm T}, x_1, 0.4)$ (fourth panel, blue line). The plot of the oxygen concentration $s({\rm T},x_1,x_2)$ is displayed in the first panel, where the white, dashed line highlights the 1D cross-section corresponding to $x_2=0.4$. The red lines in the third and fourth panels highlight $\rho_{\infty}(x_1, 0.4)$ and $\mu_{\infty}(x_1, 0.4)$ computed through~\eqref{Th1iir1} and~\eqref{Th1iiir1} with $s_{\infty}(x_1, 0.4):=s({\rm T},x_1, 0.4)$ and $c_{\infty} \equiv 0$. \textit{Bottom row.} Same as the central row but for $x_2=0.2$. The space variables $x_1$ and $x_2$ are in units of $cm$, and the parameters values used are those listed in Table~\ref{Tab1}.
 }
\label{fig:4}       
\end{figure*}

\begin{figure*}[h!]
  \includegraphics[width=\textwidth]{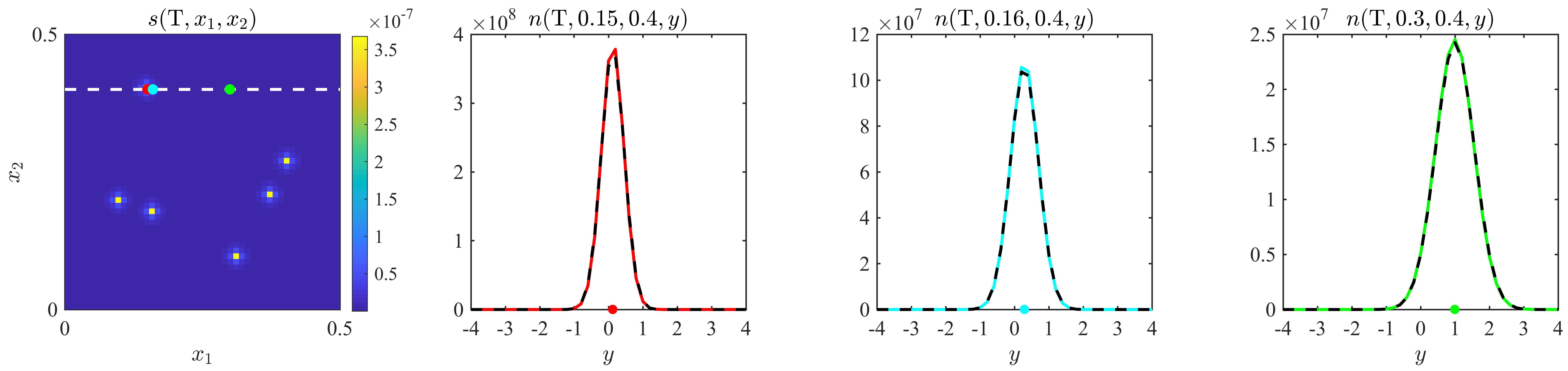}
\caption{\textbf{Two-dimensional numerical results under dynamical concentration of oxygen and in the absence of chemotherapeutic agent.} Plots of the oxygen concentration $s({\rm T},x_1,x_2)$ (first panel) and the local cell phenotypic distribution $n({\rm T},x_1,x_2,y)$ at $(x_1,x_2)=(0.15,0.4)$ (second panel), $(x_1,x_2)=(0.16,0.4)$ (third panel) and $(x_1,x_2)=(0.3,0.4)$ (fourth panel), with ${\rm T} = 5 \times 10^5 s$, obtained by solving numerically~\eqref{eq:n} and~\eqref{eq:s} imposing the initial conditions defined via~\eqref{eS1S0ic},~\eqref{num:ic1} and~\eqref{num:ic3}, the boundary condition~\eqref{eq:bcs} and assuming $c(t,x_1,x_2) \equiv 0$ (\emph{i.e.} in the absence of chemotherapeutic agent). The set $\omega$ in~\eqref{q} consists of the parts of $\Omega$ highlighted in red in the first panel of Figure~\ref{fig:4}. The white, dashed line in the first panel highlights the 1D cross-section corresponding to $x_2=0.4$ and the bullets highlight the points $(0.15,0.4)$, $(0.16,0.4)$ and $(0.3,0.4)$. In the second, third and fourth panels, the bullets on the axis of abscissas highlight the value of the local mean phenotypic state $\mu({\rm T},x_1,x_2)$ and the black, dashed lines highlight the asymptotic limit~\eqref{Th1ir1} with $\rho_{\infty}(x_1,x_2)$, $\mu_{\infty}(x_1,x_2)$ and $\sigma^2_{\infty}(x_1,x_2)$ computed through~\eqref{Th1iir1} and \eqref{Th1iiir1} with $s_{\infty}(x_1, 0.4):=s({\rm T},x_1, 0.4)$ and $c_{\infty} \equiv 0$. The space variables $x_1$ and $x_2$ are in units of $cm$, and the parameters values used are those listed in Table~\ref{Tab1}.}
\label{fig:6}       
\end{figure*}

\begin{figure*}[h!]
  \includegraphics[width=\textwidth]{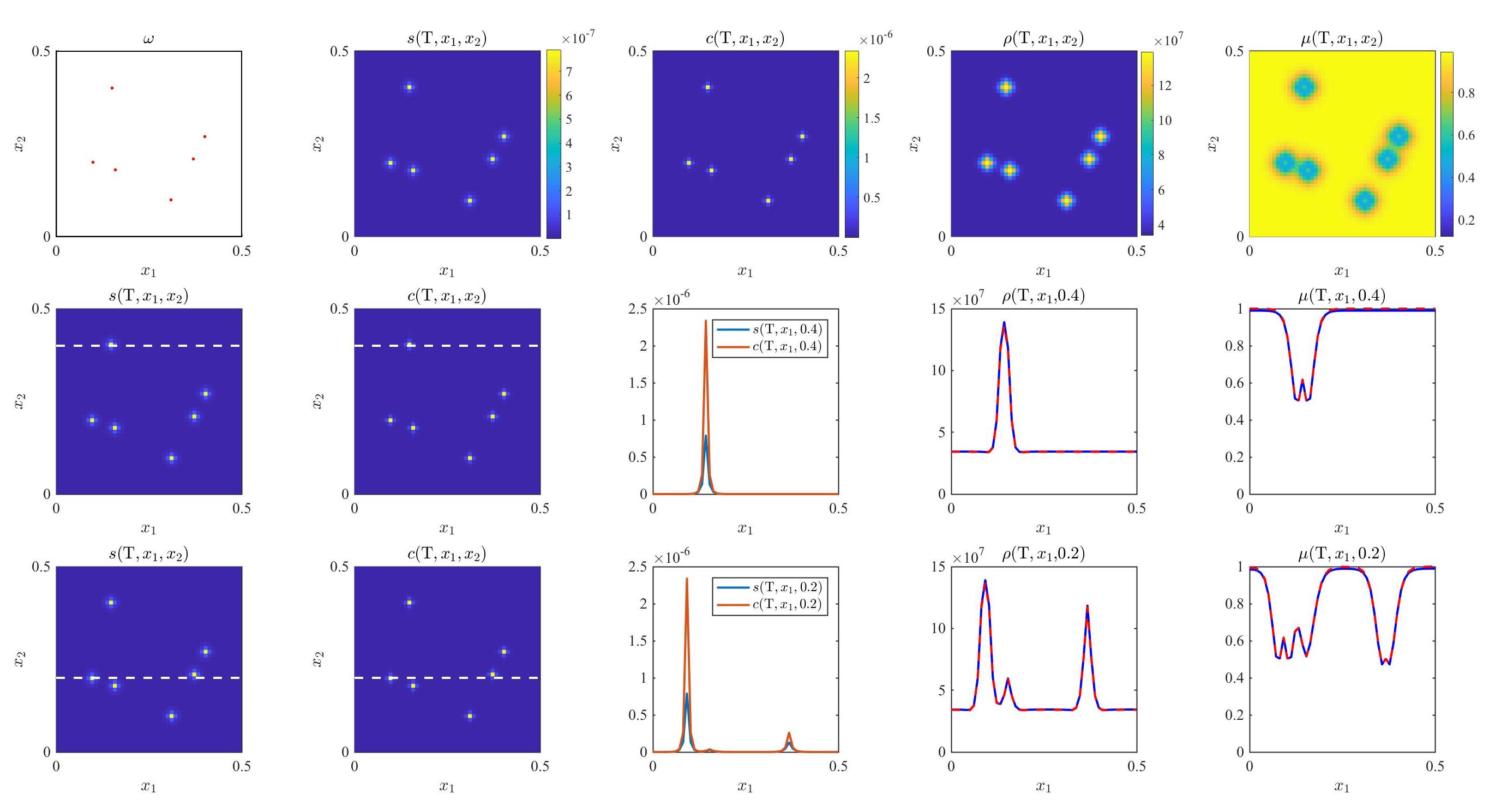}
\caption{\textbf{Two-dimensional numerical results under dynamical concentrations of oxygen and chemotherapeutic agent.} \textit{Top row.} Plots of the oxygen concentration $s({\rm T},x_1,x_2)$ (second panel), the concentration of chemotherapeutic agent $c({\rm T},x_1,x_2)$ (third panel), the cell density $\rho({\rm T},x_1,x_2)$ (fourth panel) and the local mean phenotypic state $\mu({\rm T},x_1,x_2)$ (fifth panel), with ${\rm T} = 5 \times 10^5 s$, obtained by solving numerically~\eqref{eq:n}, ~\eqref{eq:s} and~\eqref{eq:c} imposing the initial conditions defined via~\eqref{eS1S0ic},~\eqref{num:ic1} and~\eqref{num:ic3}, and the boundary conditions~\eqref{eq:bcs}. The set $\omega$ in~\eqref{q} consists of the parts of $\Omega$ highlighted in red in the first panel. \textit{Central row.} Plots of the oxygen concentration $s({\rm T},x_1, 0.4)$ (third panel, blue line), the concentration of chemotherapeutic agent $c({\rm T},x_1,0.4)$ (third panel, orange line), the cell density $\rho({\rm T}, x_1, 0.4)$ (fourth panel, blue line) and the local mean phenotypic state $\mu({\rm T}, x_1, 0.4)$ (fifth panel, blue line). The plots of the oxygen concentration $s({\rm T},x_1,x_2)$ and the concentration of chemotherapeutic agent $c({\rm T},x_1,x_2)$ are displayed in the first and second panels, where the white, dashed lines highlight the 1D cross-section corresponding to $x_2=0.4$. The red lines in the fourth and fifth panels highlight $\rho_{\infty}(x_1, 0.4)$ and $\mu_{\infty}(x_1, 0.4)$ computed through~\eqref{Th1iir1} and~\eqref{Th1iiir1} with $s_{\infty}(x_1, 0.4):=s({\rm T},x_1, 0.4)$ and $c_{\infty}(x_1, 0.4) := c({\rm T},x_1, 0.4)$. \textit{Bottom row.} Same as the central row but for $x_2=0.2$. The space variables $x_1$ and $x_2$ are in units of $cm$, and the parameters values used are those listed in Table~\ref{Tab1}.}
\label{fig:5}       
\end{figure*}

\begin{figure*}[h!]
  \includegraphics[width=\textwidth]{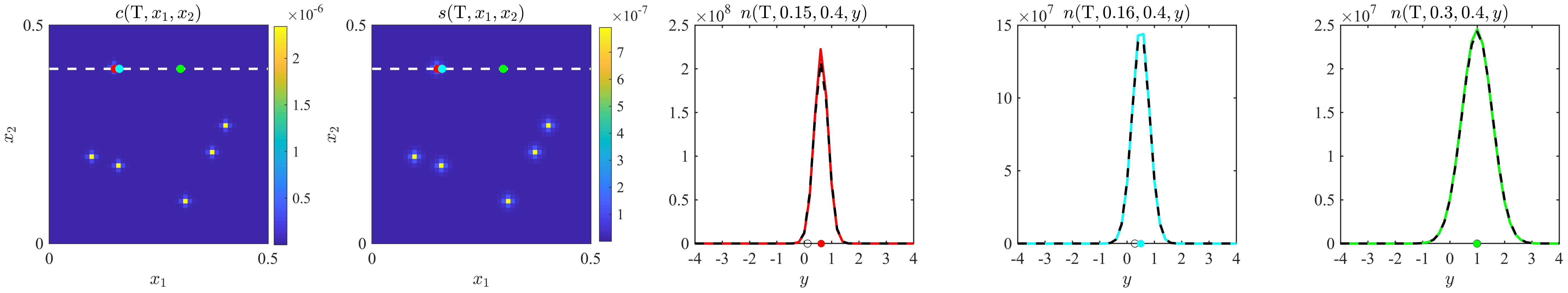}
\caption{\textbf{Two-dimensional numerical results under dynamical concentrations of oxygen and chemotherapeutic agent.} Plots of the oxygen concentration $s({\rm T},x_1,x_2)$ (first panel), the concentration of chemotherapeutic agent $c({\rm T},x_1,x_2)$ (second panel) and the local phenotypic cell distribution $n({\rm T},x_1,x_2,y)$ at $(x_1,x_2)=(0.15,0.4)$ (third panel), $(x_1,x_2)=(0.16,0.4)$ (fourth panel) and $(x_1,x_2)=(0.3,0.4)$ (fifth panel), with ${\rm T} = 5 \times 10^5 s$, obtained by solving numerically~\eqref{eq:n}, ~\eqref{eq:s} and~\eqref{eq:c} imposing the initial conditions defined via~\eqref{eS1S0ic},~\eqref{num:ic1} and~\eqref{num:ic3}, and the boundary conditions~\eqref{eq:bcs}. The set $\omega$ in~\eqref{q} consists of the parts of $\Omega$ highlighted in red in the first panel of Figure~\ref{fig:5}. The white, dashed lines in the first and second panels highlight the 1D cross-section corresponding to $x_2=0.4$ and the bullets highlight the points $(0.15,0.4)$, $(0.16,0.4)$ and $(0.3,0.4)$. In the third, fourth and fifth panels, the filled bullets on the axis of abscissas highlight the value of the mean phenotypic state $\mu({\rm T},x_1,x_2)$, while the empty bullets highlight the corresponding values obtained in the case where $c(t,x_1,x_2) \equiv 0$ (\emph{i.e.} in the absence of chemotherapeutic agent). Moreover, the black, dashed lines highlight the asymptotic limit~\eqref{Th1ir1} with $\rho_{\infty}({\bf x})$, $\mu_{\infty}({\bf x})$ and $\sigma^2_{\infty}({\bf x})$ computed through~\eqref{Th1iir1} and \eqref{Th1iiir1} with $s_{\infty}(x_1, 0.4):=s({\rm T},x_1, 0.4)$ and $c_{\infty}(x_1, 0.4) := c({\rm T},x_1, 0.4)$. The space variables $x_1$ and $x_2$ are in units of $cm$, and the parameters values used are those listed in Table~\ref{Tab1}.}
\label{fig:7}       
\end{figure*}

In the remainder of this section, we use the notation ${\bf x} \equiv (x_1,x_2)$. The sample of numerical results presented in Figure~\ref{fig:4} and Figure~\ref{fig:6} refer to the case where the oxygen concentration $s(t,x_1,x_2)$ is governed by~\eqref{eq:s}, subject to the initial condition~\eqref{num:ic3} and the boundary condition~\eqref{eq:bcs}, while the concentration of chemotherapeutic agent $c(t,x_1,x_2) \equiv 0$. On the other hand, the results presented in Figure~\ref{fig:5} and Figure~\ref{fig:7} refer to the case where $s(t,x_1,x_2)$ and $c(t,x_1,x_2)$ are governed by~\eqref{eq:s} and~\eqref{eq:c}, respectively, subject to the initial conditions~\eqref{num:ic3} and the boundary conditions~\eqref{eq:bcs}. In both cases, the set of points within the tumour tissue which are occupied by blood vessels (\emph{i.e.} the set $\omega$) is defined as illustrated by the plots in the first panels of Figure~\ref{fig:4} and Figure~\ref{fig:5}.

\paragraph{\bf Agreement between analytical and numerical results.} The sample of numerical results presented in Figure~\ref{fig:4} and Figure~\ref{fig:5} show that, in the case of constant influx from intra-tumoural blood vessels, the concentration of oxygen $s(t,x_1,x_2)$ and the concentration of chemotherapeutic agent $c(t,x_1,x_2)$ obtained by solving numerically~\eqref{eq:s} and~\eqref{eq:c}, subject to the initial conditions~\eqref{num:ic3} and the boundary conditions~\eqref{eq:bcs}, converge to some equilibria $s_{\infty}(x_1,x_2)$ and $c_{\infty}(x_1,x_2)$. As a result, in agreement with our expectation based on the results established by Theorem~\ref{Theo1} (\emph{cf.} Remark~\ref{rem1}), the cell density $\rho(t,x_1,x_2)$ and the local mean phenotypic state $\mu(t,x_1,x_2)$ computed via numerical integration of the local cell phenotypic distribution $n(t,x_1,x_2,y)$, which is obtained by solving numerically~\eqref{eq:n} subject to the initial condition defined via~\eqref{eS1S0ic} and~\eqref{num:ic1}, converge to the equilibrium values $\rho_{\infty}(x_1,x_2)$ and $\mu_{\infty}(x_1,x_2)$ given by~\eqref{Th1iir1} and~\eqref{Th1iiir1}. Moreover, the sample of numerical results presented in Figure~\ref{fig:6} and Figure~\ref{fig:7} show that the local phenotypic distribution of tumour cells $n(t,x_1,x_2,y)$ converges to the equilibrium phenotypic distribution $n_{\infty}(x_1,x_2,y)$ given by~\eqref{Th1ir1}.

\paragraph{\bf Emergence of spatial gradients of oxygen and chemotherapeutic agent.} The numerical results of Figure~\ref{fig:4} and Figure~\ref{fig:5} show that, as one would expect based on the experimental results presented by \cite{helmlinger1997interstitial}, the equilibrium concentration of oxygen $s({\rm T},x_1,x_2)$ and the equilibrium concentration of chemotherapeutic agent $c({\rm T},x_1,x_2)$ are maximal in the vicinity of the blood vessels and decrease monotonically with the distance from the blood vessels. Moreover, these results demonstrate that the nonlinear interplay between the spatial distribution of the blood vessels, the reaction-diffusion dynamics of oxygen and chemotherapeutic agent, and their consumption by tumour cells leads naturally to the emergence of spatial inhomogeneities in the equilibrium concentrations of such abiotic factors.

\paragraph{\bf Tumour cell dynamics.} The plots in Figures~\ref{fig:4}-~\ref{fig:7} demonstrate that the qualitative behaviour of the numerical results obtained under stationary concentrations of oxygen and chemotherapeutic agents displayed in Figure~\ref{fig:2} and Figure~\ref{fig:3} remains unchanged when dynamical concentrations of oxygen and chemotherapeutic agent are considered. Specifically, in the absence of chemotherapy, when moving away from the blood vessels, the equilibrium value of the cell density $\rho(t,x_1,x_2)$ decreases, the local mean phenotypic state $\mu(t,x_1,x_2)$ at equilibrium increases from values close to $y=0$ to values close to $y=1$, and the equilibrium value of the related variance $\sigma^2(t,x_1,x_2)$ increases (\emph{vid.} Figure~\ref{fig:4} and Figure~\ref{fig:6}). When chemotherapy is administered, its effect is more pronounced in the proximity of the blood vessels and consists in a reduction of the equilibrium value of $\rho(t,x_1,x_2)$, a shift of the equilibrium value of $\mu(t,x_1,x_2)$ toward $y=1$ and a reduction of the equilibrium value of $\sigma^2(t,x_1,x_2)$ compared to the case where the chemotherapeutic agent is not present. Moreover, the evolutionary dynamics of tumour cells is weakly affected by chemotherapy in regions far from the blood vessels, where the concentration of chemotherapeutic agent is lower (\emph{vid.} Figure~\ref{fig:5} and Figure~\ref{fig:7}).

\section{Conclusions and research perspectives}
\label{sec:disc}
\paragraph{{\bf Conclusions.}} The results of our analysis of evolutionary dynamics recapitulate previous theoretical results~\cite{alfarouk2013riparian,anderson2006tumor,ardavseva2019dissecting,gallaher2013evolution,gillies2012evolutionary,ibrahim2017defining,kaznatcheev2017cancer,lorenzi2018role,lorz2015modeling,marusyk2012intra,sun2015intra,villa2019modelling} and experimental data~\cite{padhani2007imaging,semenza2003targeting,sun2015intra,tannock1968relation} by demonstrating that spatial inhomogeneities in the concentration of oxygen promote the selection of different phenotypic variants at different positions within the tumour. More specifically, our analytical results indicate that the tumour tissue in the vicinity of blood vessels is to be expected to be densely populated by aerobic phenotypic variants, while poorly oxygenated regions of the tumour are more likely to be sparsely populated by anaerobic phenotypic variants. Furthermore, the analytical results obtained support the idea that higher levels of phenotypic variability may occur in hypoxic regions of the tumour, which provides a theoretical basis for experimental results such as those presented in~\cite{axelson2005hypoxia}.
\\\\
Coherently with observations made in previous theoretical and experimental studies \cite{adamski2013hypoxia,brown1998unique,powathil2012modelling,sullivan2008hypoxia,wartenberg2003regulation}, our analytical results also suggest that hypoxia favours the selection for chemoresistant phenotypic variants prior to treatment. Consequently, this facilitates the development of resistance following chemotherapy. Moreover, these results put on a rigorous mathematical basis the idea, previously suggested by formal analysis and numerical simulations ~\cite{lorenzi2018role,robertson2015impact}, that chemotherapy removes the selective barrier limiting the growth of chemoresistant phenotypic variants by killing aerobic phenotypic variants in well-oxygenated regions of the tumour.
\\\\
The results of our analysis of evolutionary dynamics are confirmed by the numerical results presented. Such numerical results also indicate that gradients of oxygen and chemotherapeutic agents, which are released from the intra-tumoural vascular network, naturally emerge in vascularised tumours due to the nonlinear interplay between the spatial distribution of the blood vessels, the reaction-diffusion dynamics of oxygen and chemotherapeutic agents, and their consumption by tumour cells.   

\paragraph{{\bf Research perspectives.}} We plan to extend our analytical results to the case where spatial movement of tumour cells is incorporated into the model. Based on the formal asymptotic results that we presented in~\cite{villa2019modelling}, in the case where cell movement is modelled through Fick's first law, we expect the qualitative behaviour of the results obtained in this paper to remain unchanged in the asymptotic regime where the rate of spontaneous phenotypic variation and the cell diffusivity tend to zero. However, further developments of the method of proof employed here are required in order to carry out a similar analysis of evolutionary dynamics in more general scenarios. 
\\\\
It would also be useful to consider a discrete version of the continuum model presented here, whereby the dynamics of tumour cells would be described in terms of a branching random walk, while the concentrations of oxygen and chemotherapeutic agent would be governed by discrete balance equations. This would make it possible to study stochastic effects which are relevant at low tumour cell densities and cannot be easily captured by deterministic models formulated in terms of differential equations. In this regard, we plan to extend the formal methods that we developed in~\cite{stace2020discrete} in order to identify the corresponding discrete counterpart of our continuum model.  
\\\\
Moreover, building upon the ideas presented in~\cite{ardavseva2019dissecting,ardavseva2020evolutionary}, it would be interesting to study the effect on the evolutionary dynamics of tumour cells of fluctuations in the rate of oxygen inflow, which are known to influence intra-tumour phenotypic heterogeneity~\cite{gillies2018eco,marusyk2012intra,robertson2015impact}. It would also be interesting to include the effect of temporal variation in the spatial distribution of intra-tumoural blood vessels, which would make it possible to explore the influence of angiogenesis on the evolutionary dynamics of tumour cells in vascularised tumours. In addition, the model considered here could be further developed to incorporate a more comprehensive description of cell metabolism that captures acidosis and enhanced tumour invasiveness caused by the presence of hypoxic cells~\cite{eales2016hypoxia,gatenby2007cellular,gatenby2007glycolysis,gillies2007hypoxia,kaznatcheev2017cancer,manem2015modeling,molavian2009fingerprint,robey2005hypoxia,zhao2013targeting}. 
\\\\
Finally, as similarly done in~\cite{almeida2019evolution} and \cite{pouchol2018asymptotic}, it would be relevant to address numerical optimal control of the model equations in order to identify possible  delivery schedules of the chemotherapeutic agent that make it possible to minimise the number of tumour cells at the end of the treatment or the average number of tumour cells during the course of treatment. In particular, it would be relevant to verify whether the results presented in~\cite{almeida2019evolution} for a spatially homogeneous model -- which indicate that continuous administration of a relatively low dose of the chemotherapy performs more closely to the optimal dosing regimen to minimise the average number of tumour cells during the course of treatment -- carry through when spatial reaction-diffusion dynamics of the chemotherapeutic agent are incorporated into the model. In this regard, it would be interesting to assess the impact of molecular properties of the chemotherapeutic agent ({\it e.g.} decay, diffusion and cellular uptake rates) and structural properties of the intra-tumoural vascular network ({\it e.g.} vascular density and blood vessels distribution) on the optimal chemotherapy schedule.

\bibliographystyle{siam}
\bibliography{bib_temp.bib}

\newpage

\appendix

\section{Proof of Proposition 1}\label{Prop1:proof}
Substituting~\eqref{def:R} and \eqref{pn2ori} into~\eqref{eq:n} yields
\begin{equation}
\label{e1-multi}
\displaystyle{\frac{\partial n}{\partial t} = \beta \frac{\partial^2 n}{\partial y^2} \; + \; \left[a - b \, (y -  h)^2 - \zeta \ \rho(t,{\bf x}) \right] n}, \quad n \equiv n(t,{\bf x},y), \quad (t,{\bf x},y) \in (0,\infty) \times \overline{\Omega} \times \mathbb{R}.
\end{equation} 
Building upon the results presented in~\cite{almeida2019evolution,chisholm2016evolutionary,lorenzi2015dissecting}, we make the ansatz~\eqref{gaussian}. Substituting this ansatz into~\eqref{e1-multi} and introducing the notation $v(t,{\bf x}) := 1/\sigma^{2}(t,{\bf x})$ we find
\beq
\label{eAnsatzTest}
\frac{\partial_t \rho}{\rho}+\frac{\partial_t v}{2v} = \frac{\partial_t v}{2}\left(y-\mu\right)^2 - \partial_t \mu \, v \left(y-\mu\right) + \,  \beta\left[v^2 \left(y-\mu\right)^2 -v \right] + \,  a  - b \, \left(y-h\right)^2 - \zeta \rho.
\eeq
Equating the second-order terms in $y$ gives the following differential equation for $v$ alone
\beq
\label{odevi}
\partial_t v + 2 \beta v^2= 2 \, b.
\eeq
Moreover, equating the coefficients of the first-order terms in $y$, and eliminating $\partial_t v$ from the resulting equation, yields
\beq
\label{odemui}
\partial_t \mu = \frac{2  b (h - \mu)}{v}.
\eeq
Lastly, choosing $y=\mu$ in~\eqref{eAnsatzTest} gives 
\begin{equation}\label{eAnzatzOneDE}
\frac{\partial_t \rho}{\rho}+\frac{\partial_t v}{2v}=-\beta v + a - b (\mu - h)^2 - \zeta \ \rho
\end{equation}
and eliminating $\partial_t v$ from~\eqref{eAnzatzOneDE} we obtain
\begin{equation} 
\label{oderhoi}
\partial_t \rho = \left[\left(a - \frac{b}{v} - b \left(\mu - h \right)^2 \right) - \zeta \rho \right]\rho.
\end{equation}
Under the initial condition~\eqref{eS1S0ic}, we have
$$
v(0,{\bf x}) = v_0({\bf x}), \quad \mu(0,{\bf x}) = \mu_0({\bf x}), \quad \rho(0,{\bf x}) = \rho_0({\bf x}),
$$
and imposing these initial conditions for~\eqref{odevi}, \eqref{odemui} and \eqref{oderhoi} we arrive at the Cauchy problem \eqref{eq:rhomuv} for the functions $v(t,{\bf x})$, $\mu(t,{\bf x})$ and $\rho(t,{\bf x})$.
\qed

\section{Proof of Theorem 1}\label{Theo1:proof}
Under assumptions \eqref{assbars} and \eqref{assbars2}, Proposition~\ref{Prop1} ensures that for any $(t,{\bf x}) \in [0,\infty) \times \overline{\Omega}$ the solution of \eqref{eq:n} subject to \eqref{eS1S0ic} and~\eqref{eS1S0icass} is of the Gaussian form \eqref{gaussian}. Therefore, building upon the method of proof presented in~\cite{ardavseva2020evolutionary,chisholm2016evolutionary}, we prove Theorem~\ref{Theo1} by studying the behaviour of the components of the solution to the Cauchy problem~\eqref{eq:rhomuv} for $t \to \infty$. 

\paragraph{Step 1: asymptotic behaviour of $v(t,{\bf x}) \equiv 1/\sigma^2(t,{\bf x})$ for $t \to \infty$.} Solving~\eqref{eq:rhomuv}$_1$ subject to the initial condition $v(0,{\bf x})=v_0({\bf x})$ gives
\begin{equation}
\label{vi}
v(t,\cdot) = \sqrt{\frac{b}{\beta}} \, \frac{\sqrt{\dfrac{b}{\beta}} + v_0 - \left(\sqrt{\dfrac{b}{\beta}} - v_0\right)\exp\left(-4\sqrt{b \, \beta} \, t\right)}
{\sqrt{\dfrac{b}{\beta}}+ v_0 + \left(\sqrt{\dfrac{b}{\beta}} - v_0\right)\exp\left(-4\sqrt{b \beta} \, t\right)},
\end{equation}  
which implies that
\begin{equation}
\label{viinf}
v(t,\cdot) \longrightarrow \sqrt{\frac{b}{\beta}} \quad \text{exponentially fast} \text{ as } t \to \infty.
\end{equation} 

\paragraph{Step 2: asymptotic behaviour of $\mu(t,{\bf x})$ for $t \to \infty$.} Solving~\eqref{eq:rhomuv}$_2$ subject to the initial condition $\mu(0,{\bf x})=\mu_0({\bf x})$ yields 
\begin{equation}
\label{mui}
\mu(t,\cdot) = \mu_0 \exp\left(-2 b\int_0^t \frac{{\rm d}z}{v(z,\cdot)}\right) + h \left[1 - \exp\left(-2b\int_0^t \frac{{\rm d}z}{v(z,\cdot)}\right) \right],
\end{equation} 
which implies that
\begin{equation}
\label{muiinf}
\mu(t,\cdot)  \longrightarrow h \quad \text{exponentially fast} \text{ as } t \to \infty.
\end{equation} 

\paragraph{Step 3: asymptotic behaviour of $\rho(t,{\bf x})$ for $t \to \infty$.} We define
$$
w(t,{\bf x}) \equiv w(v(t,{\bf x}),\mu(t,{\bf x}),S({\bf x}),C({\bf x})) := \left(\sqrt{b \, \beta} - \frac{b}{v} \right) - b \left(\mu - h \right)^2
$$
and rewrite \eqref{eq:rhomuv}$_3$ as
\begin{equation}
\label{oderhoLs0}
\partial_t \rho = \left[\left(a - \sqrt{b \, \beta} + w\right) - \zeta \rho\right] \rho.
\end{equation}
Solving~\eqref{oderhoLs0} subject to the initial condition $\rho(0,{\bf x})=\rho_0({\bf x})$ yields
\begin{equation}
\label{rhoLs0}
\rho(t,\cdot) = \frac{\displaystyle \rho_0 \exp\left[\left(a - \sqrt{b \, \beta}\right) t + \int_0^t  w(z,\cdot) \, {\rm d}z\right]}{\displaystyle 1 + \zeta \, \rho_0 \int_0^t \exp\left[\left(a - \sqrt{b \, \beta}\right) z + \int_0^z w(\tau,\cdot) \, {\rm d}\tau\right]{\rm d}z}.
\end{equation}
The asymptotic results~\eqref{viinf} and \eqref{muiinf} ensure that
\begin{equation}
\label{newresetatozero}
w(t,\cdot)  \longrightarrow 0 \quad \text{exponentially fast} \text{ as } t \to \infty ,
\end{equation}
and, therefore, \eqref{rhoLs0} allows us to conclude that
\beq
\label{rhto0}
\text{if } \sqrt{b(S({\bf x}),C({\bf x})) \, \beta} \geq a(S({\bf x}),C({\bf x})) \; \text{ then } \; \rho(t,{\bf x}) \longrightarrow 0 \; \text{ as } t \to \infty.
\eeq
On the other hand, the asymptotic result~\eqref{newresetatozero} implies that in the asymptotic regime $t \to \infty$ we have
$$
\exp\left[\left(a - \sqrt{b \, \beta}\right) t + \int_0^t  w(z,\cdot) \, {\rm d}z\right] \sim A(S,C) \, \exp\left[\left(a - \sqrt{b \, \beta}\right) t \right],
$$
and also that, under the additional assumption $\sqrt{b \, \beta} < a$, 
$$
\int_0^t \exp\left[\left(a - \sqrt{b \, \beta}\right) z + \int_0^z w(\tau,\cdot) \, {\rm d}\tau\right]{\rm d}z \sim
A(S,C) \, \frac{\exp\left[{\left(a - \sqrt{b \, \beta}\right) \, t}\right]}{a - \sqrt{b \, \beta}},
$$
for some positive function $A(S,C)$. These asymptotic relations, along with~\eqref{rhoLs0}, allow us to conclude that
\beq
\label{rhtogr0}
\text{if } \sqrt{b(S({\bf x}),C({\bf x})) \, \beta} < a(S({\bf x}),C({\bf x})) \; \text{ then } \; \rho(t,{\bf x}) \longrightarrow \frac{a(S({\bf x}),C({\bf x})) - \sqrt{b(S({\bf x}),C({\bf x})) \, \beta}}{\zeta} \; \text{ as } t \to \infty.
\eeq
Taken together, the asymptotic results~\eqref{rhto0} and \eqref{rhtogr0} ensure that
\beq
\label{rhoinf}
\rho(t,\cdot)  \longrightarrow \max\left(0, \frac{a - \sqrt{b \, \beta}}{\zeta}\right) \quad \text{as } t \to \infty.
\eeq
Claims~\eqref{Th1i}-\eqref{Th1ii} follow from the asymptotic results~\eqref{viinf}, \eqref{muiinf} and \eqref{rhoinf}.
\qed

\end{document}